\documentclass[aps,prb,twocolumn,showpacs,floatfix,10pt]{revtex4-1}

\usepackage[caption=false]{subfig}
\usepackage{verbatim}
\usepackage{graphicx}
\usepackage{dcolumn}
\usepackage{bm}
\usepackage{color}
\usepackage[colorlinks=true, allcolors=blue]{hyperref}
\usepackage{makeidx}
\usepackage{amsmath}
\usepackage{amssymb}
\usepackage{mathtools}
\usepackage{braket}
\usepackage{mathrsfs}
\usepackage{soul} 
\makeindex

\def\trace{\text{Tr}}

\DeclareMathOperator{\dif}{\mathrm{d}\!}
\renewcommand{\epsilon}{\varepsilon}

\begin{document}
	\title{Charge and energy fractionalization {mechanism} in one-dimensional channels}
	\author{Matteo Acciai$^{1,2}$, Alessio Calzona$^{1,2,3}$, Giacomo Dolcetto$^{3}$, Thomas L.\ Schmidt$^{3}$, and Maura Sassetti$^{1,2}$}
	\affiliation{ $^1$ Dipartimento di Fisica, Universit\`a di Genova, Via Dodecaneso 33, 16146, Genova, Italy\\
		$^2$ SPIN-CNR, Via Dodecaneso 33, 16146, Genova, Italy\\
		$^3$ Physics and Materials Science Research Unit, University of Luxembourg, L-1511 Luxembourg}
	\date{\today}
	\begin{abstract}
We study the problem of injecting {single} electrons into interacting
one-dimensional quantum systems{, a fundamental building block for electron quantum optics}. It is well known that such injection
leads to charge and energy fractionalization. We elucidate this concept by calculating the nonequilibrium electron distribution
function in the momentum and energy domains after the injection {of an energy resolved electron.} Our results shed light on how fractionalization occurs via the creation of
particle-hole pairs by the injected electron. {In particular we focus on systems with a pair of counterpropagating channels and we fully analyze the properties of each chiral fractional excitation which is created by the injection}. We suggest possible routes to access {their energy and momentum} distribution functions in topological quantum Hall or quantum spin Hall edge states.
	\end{abstract}
	\maketitle
	\section{Introduction}
	The past decade has witnessed a very fast development of the research field known as electron quantum optics.\cite{bocquillon2014electron,grenier2011electronoptics} Its aim is to prepare, manipulate, and measure coherent single-electron excitations, in close analogy to photon quantum optics. Pioneering experiments studied the electronic analog of the Hanbury Brown and Twiss geometry\cite{henny99-hbt,oliver99-hbt} and the Mach-Zehnder interferometer,\cite{jin03-mz} using stationary sources based on voltage-biased contacts. A major breakthrough was the experimental implementation of an on-demand single-electron source by F\`eve \emph{et al.}.\cite{feve2007mesoscopic,mahe2010mesoscopic} They showed that a periodically driven mesoscopic capacitor\cite{buttiker1993mesoscopic,moskalet2008mesoscopic} can coherently inject, for each period of the drive, a single electron and a single hole into a two-dimensional electron gas. A different kind of single-electron source was theoretically investigated by Levitov \emph{et al.},\cite{levitov96-97-06} who showed that Lorentzian voltage pulses applied to quantum conductors generate clean single-electron excitations\cite{rech2017}, without additional particle-hole creation. This prediction was also experimentally verified recently.\cite{dubois2013levitonsNature} 

The other two key ingredients necessary to perform electron quantum optics experiments are phase-coherent waveguides for electrons and beam splitters. Concerning the former, one-dimensional (1D) ballistic channels are an ideal framework. The chiral edge channels of the integer quantum Hall effect and the helical edge channels of two-dimensional topological insulators (2DTIs)\cite{hasan2010colloquium,qi2011topological,bhz2006,dolcettoreview,konig2007quantum,liu2008inasgasb,lingjie2015inasgasb,knez2011inasgasb} are notable examples. Concerning the latter, the creation of quantum point contacts in these systems enables the partition and recombination of the incoming fluxes, thus realizing the electronic version of a beam splitter.
	
	The physics of 1D systems is a fascinating topic. As a matter of fact, electron-electron ($e$-$e$) interactions in one dimension have very peculiar effects\cite{giamarchi2003book} and cannot be described within the Fermi liquid theory\cite{landau-fermiliquid} used to model interactions in two and three dimensions. Its 1D counterpart is Luttinger liquid theory, \cite{tomonaga1950,luttinger1963,haldane86-ll} which describes the low-energy properties of 1D interacting systems. Some of their most interesting features include spin-charge separation\cite{jompol09-spincharge,auslaender02-spincharge,schmidt09,kamata17-spincharge} and the fractionalization of charge,\cite{deshpande2010electron,barak2010interacting,maslov1995landauer,safi1995transport,pham2000fractional,garate2012noninvasive,calzona2015physe,muller17,safi1997properties,steinberg2007charge,kamata2014fractionalized,perfetto2014time,dolcetto13prb} spin\cite{calzona2015spin,pham2000fractional,das2011} and energy.\cite{karzig2011energypart,calzona16energypart,calzona2017quench-frac}
	
	After the implementation of the first single-electron sources, different electron quantum optics experiments followed\cite{bocquillon2012electron,bocquillon2013hom,jullien14,fevenatcomm,tewari16} which exploited the chiral 1D quantum Hall edge channels at filling factors $\nu=1,\,2$. In this context, the role of $e$-$e$ interactions between the copropagating channels was theoretically investigated\cite{jonckeere12hom,wahl14} and great interest has been shown in understanding interaction-induced relaxation and decoherence mechanisms after the injection of electrons in these systems.\cite{degiovanni09relaxation,lunde10relaxation,degiovanni10relaxation,levkivskj12relaxation,ferraro14decoherence,sukhorukov2016prb} To this aim the nonequilibrium momentum and energy distributions provide useful information and allow to study, for instance, how the injected energy is redistributed in the system. Recent experiments\cite{sueur10relaxation,altimiras10relaxation} reported on the measurement of the energy distribution in quantum Hall edge states using nonequilibrium spectroscopy by exploiting a tunable quantum dot as an energy filter.
	
	In addition to quantum Hall-based setups, 2DTIs have also been considered as a promising framework for electron quantum optics experiments.\cite{calzona16energypart,inhofer2013,hofer2013,ferraro14homtopo,strom2015entaglemntspin,dolcetto16entanglement,dolcini11-16} Here, two counterpropagating channels emerge on a given edge and electrons in different channels have opposite spin projection (a property known as spin-momentum locking). Moreover, elastic backscattering between the two channels is prevented by time-reversal symmetry. These peculiar properties allow for a richer phenomenology compared to quantum Hall systems whose edge states are chiral. Experimental observations of interactions in the edge channels of a 2DTI have been recently reported,\cite{Li15} indicating that the concept of Helical Luttinger Liquid\cite{Wu06} can be applied to these systems.
	
	In view of the implementations mentioned above, understanding the out-of-equilibrium properties of 1D ballistic conductors is of great interest in the context of electron quantum optics. In this paper we aim to shed light on this problem.
	We study the injection of an energy-resolved electron with fixed chirality into a pair of interacting counterpropagating channels, modeled as a Luttinger Liquid (LL), a configuration which can be experimentally realized in 2DTI- and quantum Hall-based structures, as we describe further on.
	While it has been known for many years that the fractions $(1\pm K)/2$ of the injected electron charge counterpropagate in the system, $K$ being the Luttinger parameter quantifying the interaction strength, it is not yet clear how the injected electron charge is redistributed among the possible excitations of the many-body system in the momentum space, and this is precisely the gap we want to fill. By evaluating the out of equilibrium momentum distribution, we show how fractionalization occurs via the creation of particle-hole pairs on each channel. Interestingly, they are found to be more relevant in the channel not directly coupled to the single-electron source. We show that the stronger the interaction, the more the injected electron loses its single-particle nature. This picture is further clarified by analyzing the energy distribution. It features a peak near the energy of the injected electron, which broadens as the interaction strength increases, and a relaxation tail at low energies.
	
	The paper is organized as follows. In Sec.~\ref{sec:model} we introduce the model and define the general quantities. Sec.~\ref{sec:results} contains our main results on the out of equilibrium momentum and energy distributions. Sec.~\ref{sec:conclusions} is devoted to our conclusions. Throughout the paper we set $\hbar=1$.
	
	\section{Model and setup}\label{sec:model}
\begin{figure}[t]
		\includegraphics[width=\columnwidth]{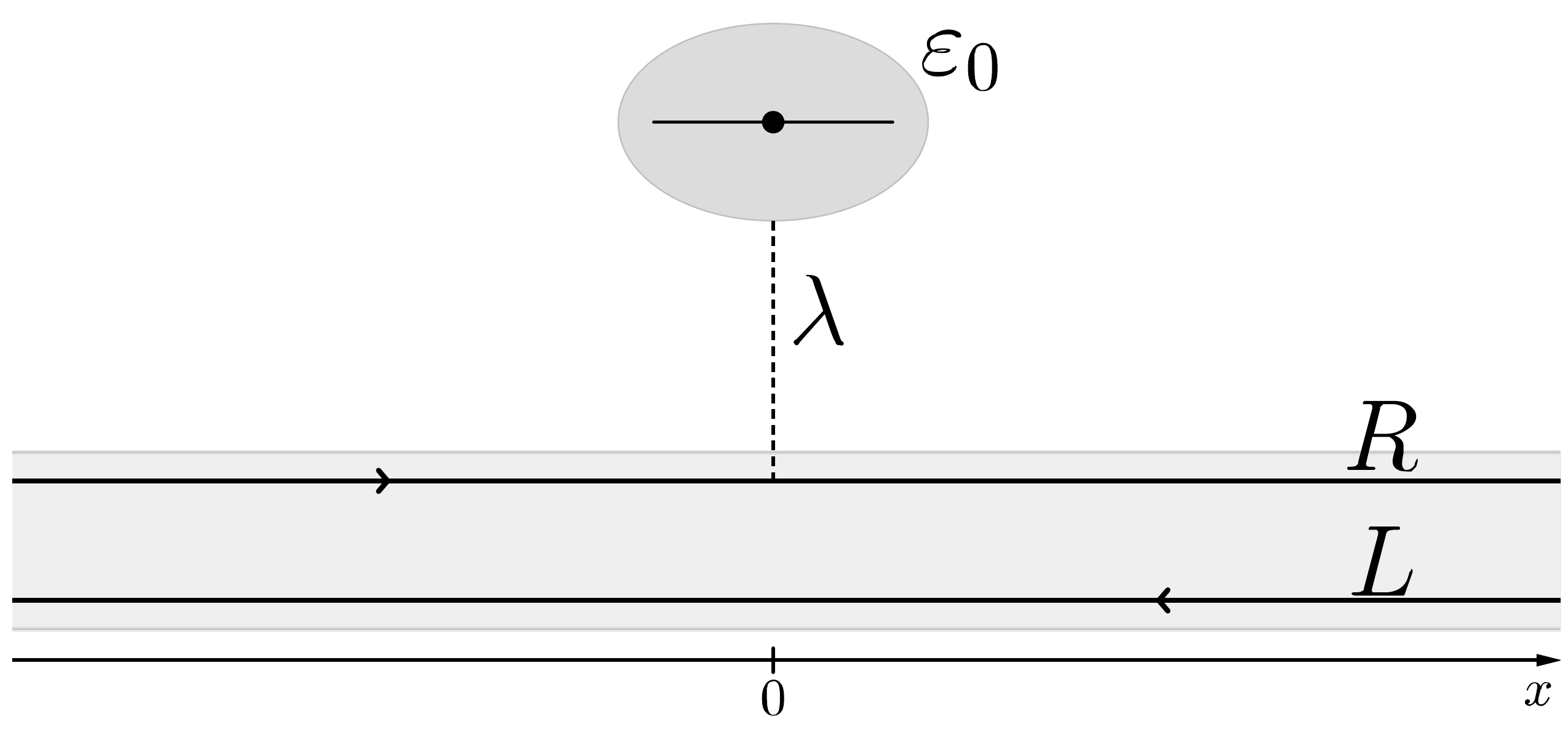}
		\caption{Sketch of the setup. A single electron with definite energy $\epsilon_0$ is locally injected from a resonant level, e.g.~a quantum dot, into the right ($R$) branch of a 1D interacting system with counterpropagating channels. The solid lines refer to the right- and left moving free-electron states ($R$/$L$ branches).}
		\label{fig:setup}
	\end{figure}

	We consider a 1D interacting system made of two counterpropagating channels. A single electron is locally injected into the 1D system from a tunnel-coupled resonant level, as sketched in Fig.~\ref{fig:setup}, which acts as a single-electron source.
	We focus on the injection of an electron on a definite branch. Such a situation can be realized in 2DTIs or in quantum Hall systems. In the former case, the 1D system is represented by the edge channels of the 2DTI and the injection of an electron with defined chirality from a mesoscopic capacitor is achieved by exploiting the spin-momentum locking.\cite{hofer2013,inhofer2013,calzona16energypart} In the latter case, the system is made of two integer quantum Hall edge states, separated by a distance such that there is appreciable $e$-$e$ interaction but negligible inter channel tunneling, as proposed in Ref.~[\onlinecite{karzig2011energypart}]. The important point is that in both situations the channels are protected from elastic backscattering.
	
	The Hamiltonian of the whole system is
	\begin{equation}
	\hat H=\hat H_{\mathrm{SL}}+\hat H_{\mathrm{LL}}+\hat H_{\mathrm{T}}\,.
	\end{equation}
	{The single level from which the injected electrons originate is modeled as 
	\begin{equation}
	\hat H_{\rm{SL}}=\epsilon_0\hat d^\dagger\hat d\,,
	\end{equation}
	with off-resonance energy $\epsilon_0>0$ measured with respect to the Fermi energy $E_{\rm{F}}$.} The 1D channels are modeled as a LL with fermion field operators $\hat{\psi}_r$ that annihilate an electron in the right ($r=R$) or left ($r=L$) branches. The Hamiltonian $\hat H_{\rm{LL}}=\hat H_0+\hat H_{\rm{int}}$ consists of a the free part
	\begin{equation}
	\hat H_0=v_{\mathrm{F}}\sum_{r=R,L}\vartheta_r\int \dif x\, \hat{\psi}_r^\dagger(x)(-i\partial_x)\hat{\psi}_r(x)\,,
	\end{equation}
	and an interaction term
	\begin{equation}
	\hat H_{\mathrm{int}}=\frac{g_4}{2}\sum_r\int\dif x\,[\hat n_r(x)]^2+g_2\int\dif x\,\hat n_R(x)\hat n_L(x)\,.
	\label{eq:int-ham}
	\end{equation}
	Here $\vartheta_{R/L}=\pm 1$, $v_\mathrm{F}$ is the Fermi velocity, $\hat n_r=\!\hat{\psi}_r^\dagger\hat{\psi}_r$ is the $r$-branch particle density and the coupling constants $g_2$ and $g_4$ refer to the inter- and intra-branch interaction respectively.
	
	The standard bosonization procedure \cite{vondelft} allows us to express the fermion fields through bosonic operators $\hat{\phi}_r$
	\begin{equation}
	\hat{\psi}_r(x)=\frac{1}{\sqrt{2\pi a}}e^{i\vartheta_rk_{\mathrm{F}}x}e^{-i\sqrt{2\pi}\hat{\phi}_r(x)}\,,
	\label{eq:bosonization}
	\end{equation}
	where $a$ is a short distance cutoff and $k_{\rm F}$ the Fermi momentum. The interacting Hamiltonian $\hat{H}_{\rm LL}$ can then be cast into a bosonic diagonal form
	\begin{equation}
	\hat H_{\mathrm{LL}}=\frac{u}{2}\sum_{\eta=\pm}\int\dif x\left[\partial_x\hat{\phi}_\eta(x)\right]^2
	\end{equation}
	with $u=(2\pi)^{-1}\sqrt{(2\pi v_{\mathrm{F}}+g_4)^2+{g_2}^2}$ the renormalized velocity and $\hat{\phi}_\eta$  chiral boson fields, i.e. $\hat{\phi}_\eta(x,t)=\hat{\phi}_\eta(x-\eta ut)$. They are related to $\hat{\phi}_r$ via
	\begin{align}
	\hat{\phi}_r(x)&=\sum_{\eta=\pm}A_{(\vartheta_r\eta)}\hat{\phi}_\eta(x),\notag \\
    A_{\pm}&=\frac{1}{2}\left(\frac{1}{\sqrt{K}}\pm\sqrt{K}\right),\,
	\label{eq:chiral-bosons}
	\end{align}
	where
	\begin{equation}
	\label{eq:KLutt}
	K=\sqrt{\frac{2\pi v_{\mathrm{F}}+g_4-g_2}{2\pi v_{\mathrm{F}}+g_4+g_2}}\,
	\end{equation}
	is the Luttinger parameter \cite{voit,miranda,kleimann02} which, for repulsive interactions, is bounded between $0<K<1$. In the non-interacting case one has $K=1$ and thus $\hat \phi_\pm=\hat \phi_{R/L}$. {It is worth noting that here the Luttinger parameter is a free parameter, allowing us to investigate a wide range of interaction strengths. By contrast, electron injection in co-propagating quantum Hall channels is usually studied only in the strong-coupling regime,\cite{wahl14,sukhorukov2016prb} the experimentally relevant one in that kind of systems.}
	Finally, the local tunneling of a $R$-branch electron is
	\begin{equation}
	\hat H_{\mathrm{T}}(t)=\theta(t)\left[\lambda\hat{\psi}_R^\dagger(0)\hat d+\text{H.c.}\right]\,,
	\end{equation}
	where $\lambda$ is a weak tunneling amplitude and $\theta$ is the Heaviside step function.
	
	{Note that we neglect Coulomb interaction contributions between the single level and the edge states. This is a rather good approximation in describing the single electron injection from a mesoscopic capacitor  because of the screening effects of the top gate.\cite{feve2007mesoscopic,grenier2011electronoptics,fevenatcomm} 
	However, it is worth pointing out that this is in general not true for other physical systems, where the Coulomb interaction between a dot and a 1D lead can have relevant effects.\cite{chernii14,elste10}}

	\section{Nonequilibrium dynamics}\label{sec:results}
	Being tunnel coupled with the LL, the resonant level acquires a finite lifetime\cite{feve2007mesoscopic,bocquillon2013hom,iyoda2014} $(2\gamma)^{-1}$. {In an interacting system, $\gamma$ depends on the interaction strength and, at lowest order in the tunneling, it reads\cite{calzona16energypart}}
			\begin{equation}
			\gamma = \frac{|\lambda|^2}{2 u} \left(\frac{a \epsilon_0}{u}\right)^{2A_-^2} \frac{e^{-\frac{a \epsilon_0}{u}}}{\Gamma(1+2A_-^2)}\,.
			\end{equation}
{	Note that, as long as $|\lambda|^2\ll u\epsilon_0$, the single level is well defined in energy, i.e. $0<\gamma \ll \epsilon_0$. In the following we will focus on this regime, known as the ``optimal regime'' in the electron quantum optics  community\cite{bocquillon2014electron,grenier2011electronoptics}, where the assumption of a single electron injection holds. 
	In order to describe the discharging of the single level, we explicitly take into account its large but finite lifetime via the approximate correlator\cite{note_correlator,calzona16energypart}}
	\begin{equation}
	\label{eq:dot_correlator}
	\braket{\hat d^\dagger(t_1)\hat d(t_2)}=e^{i\epsilon_0(t_1-t_2)}e^{-\gamma(t_1+t_2)}\,.
	\end{equation}
{	Eq.\ \eqref{eq:dot_correlator} consists in a Markov approximation, already exploited in literature\cite{sukhorukov2016prb,elste10}}, {and guarantees the conservation of the total injected charge\cite{calzona16energypart}. Here, we have not considered energy-dependent corrections to the self energy of the single-level\cite{lerner08,wachter07} since, in the optimal regime, their effects can be neglected.}

	All the other averages will be calculated using a perturbative approach in the tunneling; the time evolution of the operators will be computed in the interaction picture with respect to $\hat H_{\mathrm{T}}$. Denoting by $\hat\varrho(t)$ the time-dependent density matrix of the whole system, the average variation of an operator $\hat O(t)$, induced by the tunneling process, is defined as
	\begin{equation}
	\delta O(t)=\trace\{[\hat{\varrho}(t)-\hat{\varrho}(0)]\hat O(t)\}\,,
	\end{equation}
	where $\hat\varrho(0)$ is the equilibrium density matrix of the initial state corresponding to the LL ground state $\ket{\Omega}$ and the occupied resonant level: $\hat{\varrho}(0)=\ket{\Omega} {\bra{\Omega}}\otimes\ket{1}\bra{1}$. To lowest order in tunneling one has \cite{calzona16energypart}
	\begin{align}
	\delta O(t)&=\int_0^t\dif t_1\int_0^{t_1}\dif t_2\,\trace\{\hat{\varrho}(0) \hat H_{\mathrm{T}}(t_2)[\hat O(t),\hat H_{\mathrm{T}}(t_1)]\}\notag\\
	&+\int_0^t\dif t_1\int_0^{t_1}\dif t_2\,\trace\{ \hat{\varrho}(0) \hat H_{\mathrm{T}}(t_2)[\hat O^\dagger(t),\hat H_{\mathrm{T}}(t_1)]\}^*\,.
	\label{eq:general-average}
	\end{align}
	
	\subsection{Single-electron coherence}\label{subsec:spc}
	The coherence properties at single-particle level are described by the single-electron coherence correlator\cite{grenier2011electronoptics,grenier11,bocquillon2014electron}
	\begin{equation}
	{\mathscr{G}}_r(s,t;\xi,z)=\Braket{\hat{\psi}_r^\dagger\left(s-\frac{\xi}{2},t-\frac{z}{2}\right)\hat{\psi}_r\left(s+\frac{\xi}{2},t+\frac{z}{2}\right)}\,,
	\label{eq:spc}
	\end{equation}
	in analogy with Glauber's optical coherence.\cite{glauber62-63} Various properties of the system can be obtained from it, for instance the particle density profile\cite{calzona16energypart} and, as we will see, the electron momentum and energy distributions.
	Despite a close parallelism between electronic many-body systems and quantum optics, there are also important differences, one of them being that, even at equilibrium, the single-electron coherence does not vanish because of the presence of the Fermi sea. For this reason it is a standard procedure\cite{grenier11,grenier2011electronoptics,bocquillon2014electron} to focus on its deviations  from the equilibrium value $\mathscr{G}_r^{0}$, and thus to consider $\delta\mathscr{G}_r=\mathscr{G}_r-\mathscr{G}_r^{0}$.
	Using Eq.\ \eqref{eq:general-average}, we find the structure
	\begin{equation}
	\delta \mathscr{G}_r(s,t;\xi,z)=g_r(s,t;\xi,z)+g_r^*(s,t;-\xi,-z)\,.
	\label{eq:g-gstar}
	\end{equation}
	The detailed evaluation of functions $g_r$ is shown in Appendix~\ref{app:spc}, while here we focus on their dependence on space and time variables. First of all, it is clear that the $s$ and $t$ dependence must be retained since the system is not invariant under either space or time translations because of the injection process. Moreover, in presence of interactions the electron injected in the $R$-branch fractionalizes into two counterpropagating chiral excitations. In the long-time limit $t\gg\gamma^{-1}$, i.e. when the injection is over, they are spatially separated and they contribute independently to the single-electron coherence correlator. In Appendix~\ref{app:spc}, we demonstrate that this is indeed the case: the functions $g_r$ can be written as the sum of two chiral terms $g_{r,+}$ and $g_{r,-}$, the former propagating to the right and the latter to the left
	\begin{equation}
	g_r(s,t;\xi,z)=\sum_{\eta=\pm}g_{r,\eta}(s-\eta ut;\xi,z)\,.
	\label{eq:g-sum}
	\end{equation}
	This important relation allows us to separately study the dynamical properties of the two chiral fractional excitations. We find the following expression for functions $g_{r,\eta}$
	\begin{equation}
	\begin{split}
	&g_{r,\eta}(x_\eta;\xi,z)=\frac{|\lambda|^2}{(2\pi a)^2}\,\mathcal{C}_{r,\eta}(\xi,z)e^{i\vartheta_r k_{\mathrm{F}}\xi}\\
	&\quad\times\iint_0^{+\infty}\dif t_1\dif \tau\, e^{-2\gamma t_1}\mathcal{F}(\tau)\Xi_{r,\eta}(x_\eta,\zeta_\eta,t_1,\tau)\,,
	\end{split}
	\label{eq:g-r_eta-result}
	\end{equation}
	where $x_\eta=s-\eta ut$, $\zeta_\eta=\xi-\eta uz$ and
	\begin{subequations}
		\begin{equation}
		\begin{split}
		\mathcal{C}_{r,\eta}(\xi,z)&=\left[\frac{a}{a-iuz+i\vartheta_r\xi}\right]^{A_+^2}\\ & \times \left[\frac{a}{a-iuz-i\vartheta_r\xi}\right]^{A_-^2}\; \frac{\zeta_\eta-i a \vartheta_r}{\zeta_\eta}\,,
		\end{split}
		\label{eq:g-xi-z}
		\end{equation}
		\begin{equation}
		\mathcal{F}(\tau)=e^{-\gamma\tau-i\epsilon_0\tau}\left[\frac{a}{a-iu\tau}\right]^{1+2A_-^2}\,,
		\end{equation}
		\begin{equation}
		\begin{split}
		\label{eq:Xi}
		&\Xi_{r,\eta}(x_\eta,\zeta_\eta,t_1,\tau)=\left[\frac{a+i\eta(x_\eta+\zeta_\eta/2+\eta ut_1)}{a+i\eta(x_\eta-\zeta_\eta/2+\eta ut_1)}\right]^{\alpha_{r,\eta}}\\
		&\times 2i\,\text{Im}\left\{\left[\frac{a-i\eta(x_\eta-\zeta_\eta/2+\eta u(t_1+\tau))}{a-i\eta(x_\eta+\zeta_\eta/2+\eta u(t_1+\tau))}\right]^{\alpha_{r,\eta}}\right\}\,.
		\end{split}
		\end{equation}
	\end{subequations}
	Here, we assumed that the thermal energy is much smaller than the typical excitation energies of the system and we thus considered $T\to0$. The exponents $\alpha_{r,\eta}$ are related to the $A_\eta$ coefficients in Eq. \eqref{eq:chiral-bosons} by
	\begin{equation}
	\label{eq:exponents}
	\alpha_{R,\eta}=A_\eta^2 \, , \quad   \alpha_{L,\eta}=A_+A_-\,. 
	\end{equation}
	 The factor $(\zeta_\eta-i a \vartheta_r)\zeta_\eta^{-1}$ in Eq.~\eqref{eq:g-xi-z} stems from the point-splitting procedure\cite{vondelft, note_pointsplitting} and ensures that the diagonal part of the single-electron coherence truly represents the electron particle density $\delta \rho_r(s,t)=\delta \mathscr{G}_r(s,t,;0,0)$ (see Appendix \ref{app:ps} for details). The previous expressions will be the building blocks from which the energy and momentum distributions can be obtained.

	\subsection{Momentum distribution}
	The momentum distribution of the $R$ and $L$ branches is defined as the average variation [as in Eq.~\eqref{eq:general-average}] of the occupation number operator
	\begin{equation}
	\hat n_r(k,t)=\hat c^\dagger_{r,k}(t)\hat c_{r,k}(t)\,,
	\end{equation}
	where $\hat c_{r,k}$ annihilates an electron with momentum $k$ on the $r$-branch ($r = R,L$). Using the single-electron coherence, one can represent the occupation number variation as
	\begin{equation}
	\label{eq:nk_step1}
	\delta n_r(k,t)=\frac{1}{2\pi}\iint_{-\infty}^{+\infty}\dif \xi\,\dif s\, e^{-ik\xi}\delta\mathscr{G}_r(s,t;\xi,0)\,.
	\end{equation}
	{In general, the momentum distribution $\delta n_r{(k,t)}$ has a temporal evolution \cite{ferraro14decoherence}. Focusing on the long-time limit $t\gg \gamma^{-1}$, however, the decoupling relation in Eq.\ \eqref{eq:g-sum} allows us to express the momentum distribution as a sum of time-independent contributions}
	\begin{equation}
	\delta n_r(k)=\sum_{\eta=\pm}\delta n_{r,\eta}(k)\,,
	\end{equation}
	where each of the four terms
	\begin{equation}
	\label{eq:nk}
	\delta n_{r,\eta}(k)=\frac{1}{\pi}\mathrm{Re}\left\{\iint_{-\infty}^{+\infty}\dif\xi\,\dif s\,e^{-ik\xi} g_{r,\eta}(s;\xi,0)\right\}
	\end{equation}
	 represents the momentum distribution of the $r$-branch electrons associated to the right ($\eta=+$) or the left ($\eta=-$) moving chiral excitation. Using Eq.~\eqref{eq:g-r_eta-result} and conveniently shifting the variable $s$, each term can be written as
	\begin{align}
	&\delta n_{r,\eta}(k)=\frac{|\lambda|^2}{2\pi\gamma}\frac{1}{(2\pi a)^2}\,\mathrm{Re}\left\{\int_0^{+\infty}\dif \tau\mathcal{F}(\tau)\right.\notag\\
	&\times\left.\int_{-\infty}^{+\infty}\dif\xi\,e^{-i(k-\vartheta_rk_{\mathrm{F}})\xi}\; \mathcal{C}_{r,\eta}(\xi,0)\,\int_{-\infty}^{+\infty}\dif s\,\chi_{r,\eta}(s,\xi,\tau)\right\},
	\label{eq:result-nk}
	\end{align}
	with
	\begin{align}
	\chi_{r,\eta}(s,\xi,\tau)&=\left[\frac{a+i\eta(s+\xi-\eta u\tau)}{a+i\eta(s-\eta u\tau)}\right]^{\alpha_{r,\eta}}\notag\\
	&\times 2i\,\mathrm{Im}\left\{\left[\frac{a-i\eta s}{a-i\eta(s+\xi)}\right]^{\alpha_{r,\eta}}\right\}\,.
	\label{eq:xi-nk}
	\end{align}
	{The time independence of the momentum distribution in the long time limit $t\gg\gamma^{-1}$ stems from the fact that our model does not take into account for spectrum non-linearities or equilibration mechanism that would induce a time evolution even on time scales greater than $\gamma^{-1}$.\cite{sukhorukov2016prb,levkivskj12relaxation}}

	\begin{figure}[tbp]
		\includegraphics[width=\columnwidth]{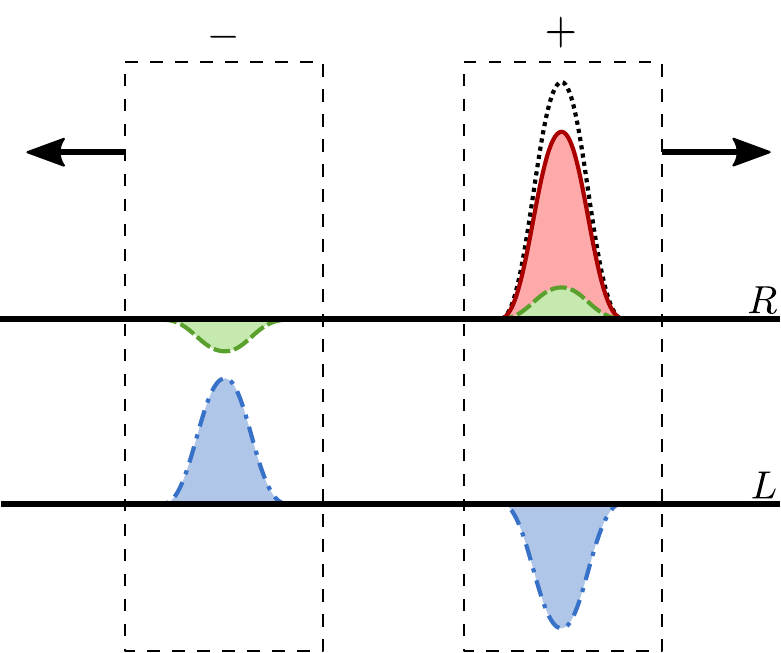}
		\caption{(color online) Sketch of the fractionalization mechanism. The real-space structure of the counterpropagating fractional excitations (highlighted by the ``$+$'' and ``$-$'' rectangles) is shown, distinguishing between the contributions from the two branches, $R$ and $L$. }
		\label{fig:cartoon}
	\end{figure}
	
	\begin{figure*}[ht]
		\includegraphics[width=2\columnwidth]{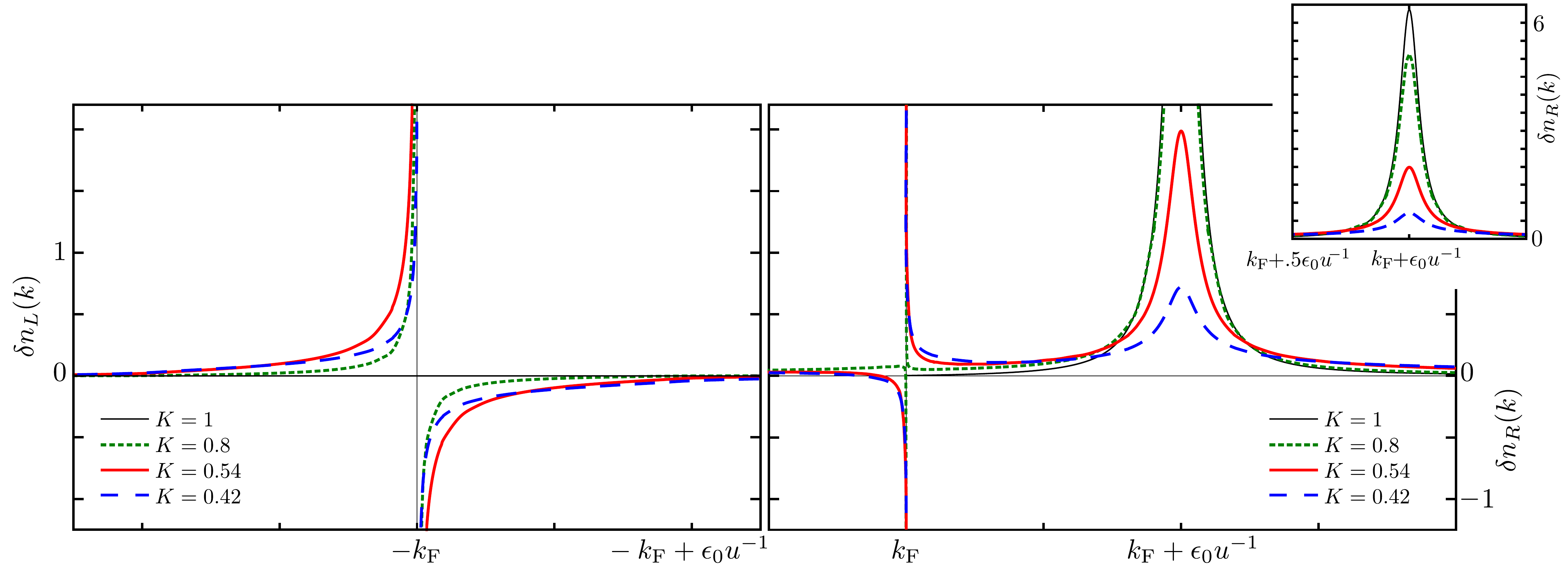}
		\caption{(color online) Momentum distribution (in units of $u\epsilon_0^{-1}$) of the $L$ branch (left panel) and $R$ branch (right panel) for different interaction strengths: $K=0.8$ (green short-dashed line), $0.54$ (red continuous line), $0.42$ (blue long-dashed line) and $1$ (thin black line). The inset focuses on the broadening of the peak centered in $k_\mathrm{F}+\epsilon_0 u^{-1}$ as the interaction strength increases. Parameters: $\epsilon_0au^{-1}=1/40$ and $\gamma=0.05\epsilon_0$.
		}
		\label{fig:nk-total}
	\end{figure*}

	In order to clarify the meaning of {the $\delta n_{r,\eta}(k)$} terms, is it useful to focus at first on the integrated quantities
	\begin{equation}\label{eq:DeltaN}
	\delta N_{r,\eta} = \int_{-\infty}^{+\infty} \delta n_{r,\eta}(k)\; dk\,,
	\end{equation}
	which represent the excess number of electrons carried by each of the two chiral excitations in the $r$ branch. A straightforward calculation leads to
	\begin{align}
	\label{eq:R+}
	\delta N_{R,+} & = 1+A_-^2 = 1 + \tfrac{1}{4} \left(K^{-1}+K-2\right)\\
	\delta N_{R,-} & = -A_-^2 = - \tfrac{1}{4} \left(K^{-1}+K-2\right)\\
	\delta N_{L,\pm} & = \mp A_+A_- = \mp \tfrac{1}{4} \left(K^{-1}-K\right)\,.
	\end{align}
	The total charge of each chiral excitation is thus 
	\begin{subequations}
		\begin{equation}
		\delta\mathcal{Q}_\eta=	e\sum_{r=R,L}\delta N_{r,\eta}= e\frac{1+\eta K}{2}\,,
		\end{equation}
		reproducing the well-known results of charge fractionalization.\cite{maslov1995landauer,safi1995transport,pham2000fractional}
		As a direct consequence of conservation of the electron number on each branch, which follows from the absence of backscattering, the following sum rules are also satisfied
		\begin{equation}
		\sum_{\eta=\pm}\delta N_{R,\eta} = 1\,,\quad \sum_{\eta=\pm}\delta N_{L,\eta} = 0\, .
		\label{eq:sumrules}
		\end{equation}
	\end{subequations}
	In Fig.~\ref{fig:cartoon} we sketch the structure of the chiral excitations in position space. The left-moving excitation is made up of a negative packet $R_-$ (in green) and a positive one $L_-$ (in blue). By contrast, the right-moving excitation is made up of a negative packet $L_+$ (in blue) and a positive one $R_+$ (dotted line). According to Eq.~\eqref{eq:R+}, the latter can be regarded as the sum of a unit packet (in red), representing the injected electron, and a positive packet (in green) with opposite charge compared to $R_-$.
	
	This scenario corresponds to the well-known fractionalization phenomenon, where the injected single-electron charge is split into counterpropagating fractional charges.
	However, being based on the integrated quantities~\eqref{eq:DeltaN}, this picture is not able to describe the detailed structure of the many-body excitations created in the 1D conductor, and these types of information are crucial to give a proper characterization of the relaxation and decoherence mechanism due to the interplay of single-electron injection and electron interaction.
	Therefore, we go beyond this coarse description in the following by characterizing the many-body nature of the fractionalization phenomenon using the momentum-resolved contributions~\eqref{eq:result-nk}.

	At first, let us consider the noninteracting case $K=1$. Here $A_+=1$, $A_-=0$ {and Eq.\ \eqref{eq:result-nk}} readily reduces to:
		\begin{align}
		\label{eq:momK1}
		\delta n_{R,+}(k) &=\theta(k-k_{\mathrm{F}})\frac{v_{\mathrm{F}}\gamma/\pi}{\gamma^2+[\epsilon_0-v_{\mathrm{F}}(k-k_{\mathrm{F}})]^2} \\ 
		\delta n_{R_-} &= \delta n_{L,\pm} = 0.\notag
	   \end{align}
For $K=1$ the right and left branches are two independent and chiral systems, so the electron injected on the $R$ branch will just propagate to the right without affecting the $L$ branch. The momentum distribution in Eq.~\eqref{eq:momK1} is a truncated Lorentzian\cite{grenier11,ferraro13wigner} of width $\gamma$, centered in $k=k_{\mathrm{F}}+\epsilon_0v_{\mathrm{F}}^{-1}$. In the limit $\gamma/\epsilon_0\to 0$ it becomes a delta function. It is worth noting that the Fermi sea remains a spectator as $\delta n_{R,+}(k)\ne 0$ only for $k>k_{\mathrm{F}}$. As we will see, this will no longer be true in presence of interactions.

	\begin{figure*}[tbp]
		\includegraphics[width=2\columnwidth]{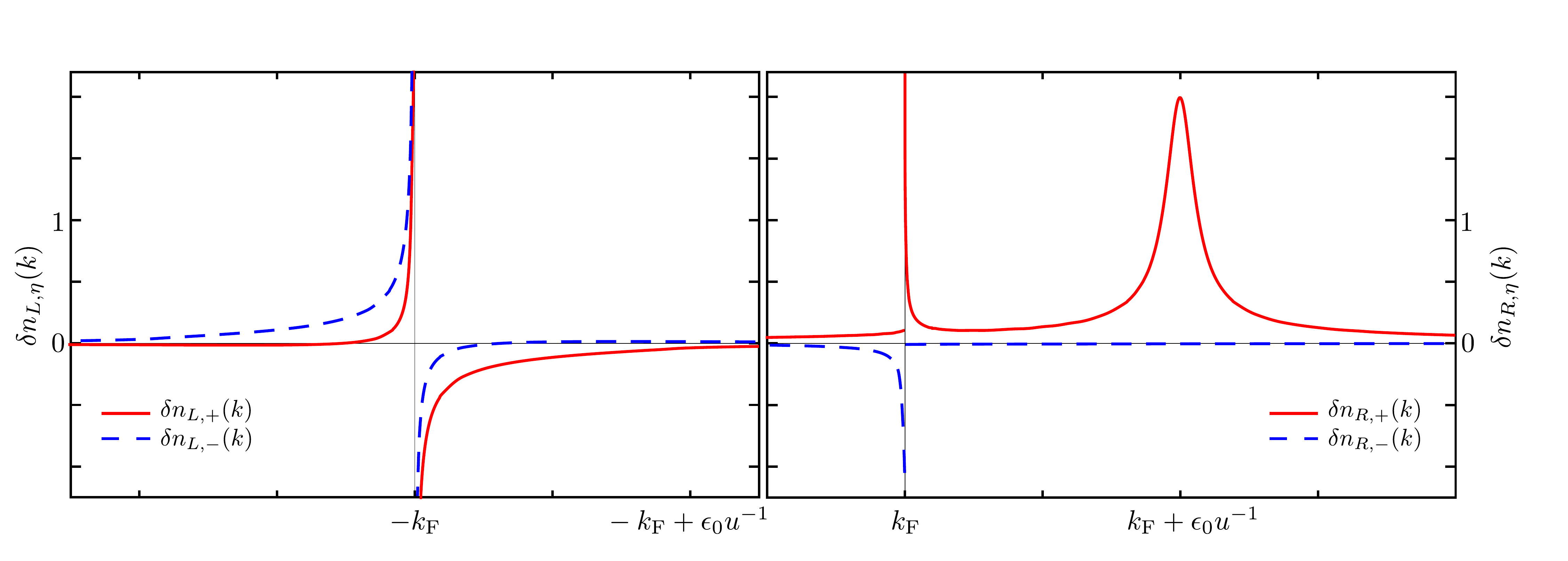}
		\caption{(color online) Chiral components of the momentum distribution (in units of $u\epsilon_0^{-1}$). The red continuous line refers to the momentum distribution of the chiral right moving excitation, i.e. $\delta n_{L,+}$ (left panel) and $\delta n_{R,+}$ (right panel). The dashed blue line refers to the momentum distribution of the chiral left moving excitation, i.e. $\delta n_{L,-}$ (left panel) and $\delta n_{R,-}$ (right panel). Parameters: $K=0.54$, $\epsilon_0au^{-1}=1/40$ and $\gamma=0.05\epsilon_0$.}
		\label{fig:nk-chiral}
	\end{figure*}
	
	In an interacting system the complete momentum distribution functions is obtained by numerically computing the integrals in Eq.~\eqref{eq:result-nk}. In Fig.~\ref{fig:nk-total} we plot $\delta n_{L}(k)$ (left panel) and $\delta n_{R}(k)$ (right panel) for different values of the interaction parameter $K$. Increasing the interaction strength, the peak around $k_{\mathrm{F}}+\epsilon_0u^{-1}$ (right panel) lowers and broadens while particle-hole contributions emerge
	around the Fermi points. In this respect, it is useful to consider the limit $k\to \pm k_{\mathrm{F}}$ where the momentum distributions $\delta n_{R/L}(k)$ exhibit a power-law behavior
	\begin{equation}
	\begin{split}
	\delta n_r({k})&\simeq\frac{u}{\pi^2 \epsilon_0}\left(\frac{{\epsilon_0 a}}{u}\right)^{2A_-^2}\Gamma(1-2A_-^2)\; C_r\\ &\quad \times \text{sgn}(k-\vartheta_r k_\mathrm{F})\,\left|u \frac{k-\vartheta_r k_\mathrm{F}}{\epsilon_0}\right|^{2 A_-^2-1}\,,
	\label{eq:scaling-nk}
	\end{split}
	\end{equation}
	with interaction-dependent coefficients
	\begin{subequations}
		\label{eq:C_r}
		\begin{align}
		C_R&=\sin(2\pi A_-^2)\sin(\pi A_-^2)
		\\C_L&=-\sin^2\!(\pi A_+A_-)\cos(\pi A_-^2)
		\,.
		\end{align}
	\end{subequations}
	Eq.~\eqref{eq:scaling-nk} is demonstrated in Appendix~\ref{app:nkscaling} and holds as long as $A_-^2<1/2$, i.e., when the interaction in not too strong ($K>0.27$). In this case, the momentum distribution features a power-law divergence at the Fermi points $\pm k_{\mathrm{F}}$. This divergence is integrable, consistently with the fact that $\delta n_r(k)$ defines a probability density, and gets weaker as the interaction strength increases. Such a behavior can be understood as a manifestation of the well-known Anderson's orthogonality catastrophe.\cite{anderson67,mahan81} {We note that, as discussed in Appendix \ref{app:nkscaling}, the exponent of the power-law behavior in Eq.\ \eqref{eq:scaling-nk} is robust with respect to the approximation made in Eq.\ \eqref{eq:dot_correlator}.}
	
	Quite interestingly, particle-hole pairs are much more relevant on the $L$ branch than on the $R$ one, as can be seen in Fig.~\ref{fig:nk-total}. This means that excitations around the Fermi points are more important on the channel which is not tunnel-coupled to the single-electron emitter.
	This  feature, which from a mathematical point of view emerges from Eqs.~\eqref{eq:C_r} where $C_L$ is greater than $C_R$, can be interpreted by relying on the following picture of the interaction mechanism.\cite{voit93spectral,takei2010spectroscopy} We note that the intra-branch coupling $g_4$ alone simply renormalizes the Fermi velocity and does not modify the Luttinger parameter $K=1$ (see Eq.~\eqref{eq:KLutt} with $g_2=0$). Therefore, it is the inter-branch coupling $g_2$ which plays a fundamental role in the fractionalization mechanism. Since the injection is performed on the right branch, a first interaction process couples the injected electrons with momentum near $\epsilon_0u^{-1}$ to the left branch, thus creating particle-hole excitations around $-k_{\mathrm{F}}$.  Then, a second process couples the excitations just created on the left branch to the right branch, exciting particle-hole pairs around $+k_{\mathrm{F}}$. The latter is thus a higher-order process compared to the creation of particle-hole pairs on the $L$ branch. For weak interactions, this heuristic picture perfectly fits with the expression of the $C_r$ coefficients. Indeed on can show that
	\begin{align}
	C_R &=  \tfrac{1}{8}\pi^2 (g_4+2\pi v_\mathrm{F})^{-4}\; (g_2)^4+ \text{O}(g_2)^6\\
	C_L &=  -\tfrac{1}{4}\pi^2 (g_4+2\pi v_\mathrm{F})^{-2}\; (g_2)^2+ \text{O}(g_2)^4.
	\end{align}
	
	Having discussed the features of $\delta n_{R}$ and $\delta n_L$ for different interaction strengths, we can now analyze the chiral components of the momentum distribution. In Fig. \ref{fig:nk-chiral}, the four terms $\delta n_{r,\eta}$ are plotted for $K=0.54$. Functions $\delta n_{L,\pm}$ are shown in the left panel, while $\delta n_{R,\pm}$ are plotted in the right one. Solid red lines refer to the chiral right-moving components ($\eta=+$) and the blue dashed ones to the chiral left-moving terms ($\eta=-$). Interestingly, it is possible to understand the features of these plots using the sketch in Fig.~\ref{fig:cartoon}. The peak on $R+$ centered around $k_\mathrm{F}+\epsilon_0 u^{-1}$ is indeed the remnant of the injected electron: it is related to the red packet in Fig. \ref{fig:cartoon}. As discussed above, the inter-branch interaction creates particle-hole pairs on the $L$-branch. However, because of the excess right-moving charge present on $R+$, the majority of the holes is ``dragged'' to the right (see the negative blue packet in Fig.~\ref{fig:cartoon}). This explains the asymmetry between $\delta n_{L+}(k)$, rich in holes and larger for $k>-k_\mathrm{F}$, and $\delta n_{L-}(k)$, rich in particles and larger for $k<-k_\mathrm{F}$. Electron-hole pairs are also created on the $R$ branch, but through a higher-order process and thus their impact on the $R$ branch is reduced. Again, the excess left-moving charge on branch $L-$, represented by the positive blue packet in Fig.~\ref{fig:cartoon}, drags the holes on the $R$ branch to the left (green negative packet) and pushes particles to the right (green positive packet). As a consequence, $\delta n_{R,-}(k)$ basically contains only holes while $\delta n_{R,+}(k)$ features a particle component near the Fermi point, superimposed on the peak tails.
	
As a last comment, we note that the momentum distribution of the right-moving excitation (solid red lines) is very different from the left-moving one (dashed blue lines): the former features the peak around $k_\mathrm{F}+\epsilon_0 u^{-1}$ while the latter has significant weight only around $-k_\mathrm{F}$. This strong asymmetry is completely lost within a real-space description of the chiral excitations. As shown in Ref.~[\onlinecite{calzona16energypart}], their particle density profile $\delta \rho_\pm(s,t)$ is in fact mirror-shaped with respect to the injection point
\begin{equation}
\frac{\delta \rho_+(s,t)}{\delta \mathcal{Q}_+} = \frac{\delta \rho_-(-s,t)}{\delta \mathcal{Q}_-}\,.
\end{equation}

We would like to stress that it is possible in principle to experimentally access every contribution $\delta n_{r,\eta}(k)$. A detector placed to the right (left) of the injection point can in fact exclusively measure the properties of the chiral right (left) moving excitation $\eta=+$ ($\eta=-$). Moreover, we observed that the interesting features of the momentum distributions are centered around the Fermi points and around $k_\mathrm{F}+\epsilon_0 u^{-1}$. Provided that $k_\mathrm{F}\gg\epsilon_0 u^{-1}$, it is thus possible to easily distinguish between the contributions from the $R$ and the $L$ branches.

\subsection{Energy distribution}
In an interacting system, energy and momentum are not related through a simple dispersion relation and are independent quantities.\cite{voit93spectral,meden1992} Therefore, the energy distribution of the excitations provides complementary information to the already discussed momentum distribution. Here, we will focus on the following component of 
the local nonequilibrium spectral function integrated over time
\begin{equation}
\label{eq:defA}
\delta \mathcal{A}_r(\omega,x_p) =  \frac{u}{2\pi} \iint_{-\infty}^{+\infty} dt\, dz \;  \delta \mathscr{G}_r(x_p,t;0,z) \; e^{i \omega z}\,.
\end{equation}
Such a quantity has the great advantage to be directly related to a physical observable, namely the total charge transferred from the system to a tunnel coupled single empty level. It can be thus experimentally accessed via quantum dot spectroscopy.\cite{sueur10relaxation, altimiras10relaxation, takei2010spectroscopy} Before explicitly computing $\delta \mathcal{A}_r$, it is worth discussing more in detail the aforementioned relation, in order to further clarify the meaning of Eq.\ \eqref{eq:defA} and to allow for a clearer interpretation of the results.

Let $\hat H_p = \omega \hat b^\dagger \hat b$ be the Hamiltonian of a probe quantum dot, modeled as a single level with energy $\omega>0$. At position $x_p$, it is tunnel coupled to the $r$-branch of the system via
\begin{equation}
\hat H_T^p = \left[ \lambda_p \hat \psi_r^\dagger(x_p) \hat b + \text{H.c.}\right]\,.
\end{equation}
The current transferred from the system to the probe dot reads
\begin{equation}
\hat I_r = i e\left[\hat H_T^p, \hat b^\dagger \hat b\right] = ie\left[\lambda_p \hat \psi_r^\dagger(x_p) \hat b -\text{H.c.}\right]
\end{equation}
and, to the lowest order in the tunneling amplitude $\lambda_p$, its average value is given by
\begin{equation}
I_r(t) = i \int_{-\infty}^{t} \left\langle \left[\hat H_\mathrm{T}^p(\tau), \hat I_r(t) \right]\right\rangle \, d\tau.
\end{equation}
We now assume that the single level is held empty, i.e. $\langle b^\dagger b \rangle=0$, considering for example an additional stronger coupling with a drain at lower chemical potential.\cite{takei2010spectroscopy} Then, the total charge transferred from the system to the dot
\begin{equation}
\begin{split}
q_r(\omega, x_p) &= \int_{-\infty}^{+\infty} I_r(t)\, dt \\
\end{split}
\end{equation}
 can then be expressed as
 \begin{equation}
q_r(\omega, x_p)= { e |\lambda_p|^2} \,\iint_{-\infty}^{\infty} dt\,dz\;  \mathscr{G}_r(x_p,t;0,z) \, e^{i \omega z}\,.
\end{equation}
The variation of this quantity, induced by the electron injection, is thus directly related to the energy distribution defined in Eq. \eqref{eq:defA} via
\begin{equation}
\delta q_r(\omega,x_p) = 2 \pi \frac{e |\lambda_p|^2}{u} \delta \mathcal{A}_r(\omega,x_p)\,.
\end{equation}
Since the energy is conserved in the tunneling process, it is clear that the function $\delta \mathcal{A}_r(\omega,x_p)$ represents the probability density of destroying an excitation with energy $\omega>0$ by extracting an electron from the $r$ branch at position $x_p$. Note that if the system is in its ground state (without the injected electron), no excitations can be destroyed and no charge can be transferred to the probe dot. As a consequence, the variation $\delta q_r$ correspond to the total transferred charge $q_r$.

If the probe dot is positioned far away from the injection point, i.e. $|x_p| \gg u \gamma^{-1}$, the chiral excitations created by the electron injection will reach it only at large time $t\gg \gamma^{-1}$. In this limit, Eq. \eqref{eq:g-sum} holds and allows to distinguish between the contributions of each chiral excitation
\begin{equation}
\delta \mathcal{A}_r(\omega,x_p) \simeq \begin{cases}
\delta \mathcal{A}_{r,+}(\omega) \qquad &x_p\gg u \gamma^{-1}\\
\delta \mathcal{A}_{r,-}(\omega) \qquad &x_p\ll - u \gamma^{-1}
\end{cases}\,.
\end{equation}
Here, the chiral energy distribution of the $r$ branch does not depend on $x_p$ and reads
\begin{equation}
\begin{split}
&\delta \mathcal{A}_{r,\eta}(\omega) =  \frac{u}{\pi}\; \text{Re}\left\{ \iint_{-\infty}^{+\infty} \!dt dz \; e^{i \omega z}\; g_{r,\eta}(ut;0,z)\right\}\\
&\quad =\frac{|\lambda|^2}{2\pi\gamma}\frac{1}{(2\pi a)^2}\,\mathrm{Re}\left\{\int_0^{+\infty}\dif \tau\mathcal{F}(\tau)\right.\\
&\left.\quad \times\int_{-\infty}^{+\infty}\dif z\,e^{i\omega z}\mathcal{C}_{r,\eta}(0,z)\int_{-\infty}^{+\infty}\dif s\,\chi_{r,\eta}(s,-\eta u z,\tau)\right\}.
\label{eq:result-energy}
\end{split}
\end{equation}
It is also possible to define a total chiral energy distribution, summing with respect to branch index $r$
\begin{equation}\label{eq:A+-}
\delta \mathcal{A}_\eta (\omega) = \sum_{r=R,L} \delta \mathcal{A}_{r,\eta}(\omega)\,.
\end{equation}

\begin{figure}[tbp]
	\includegraphics[width=\columnwidth]{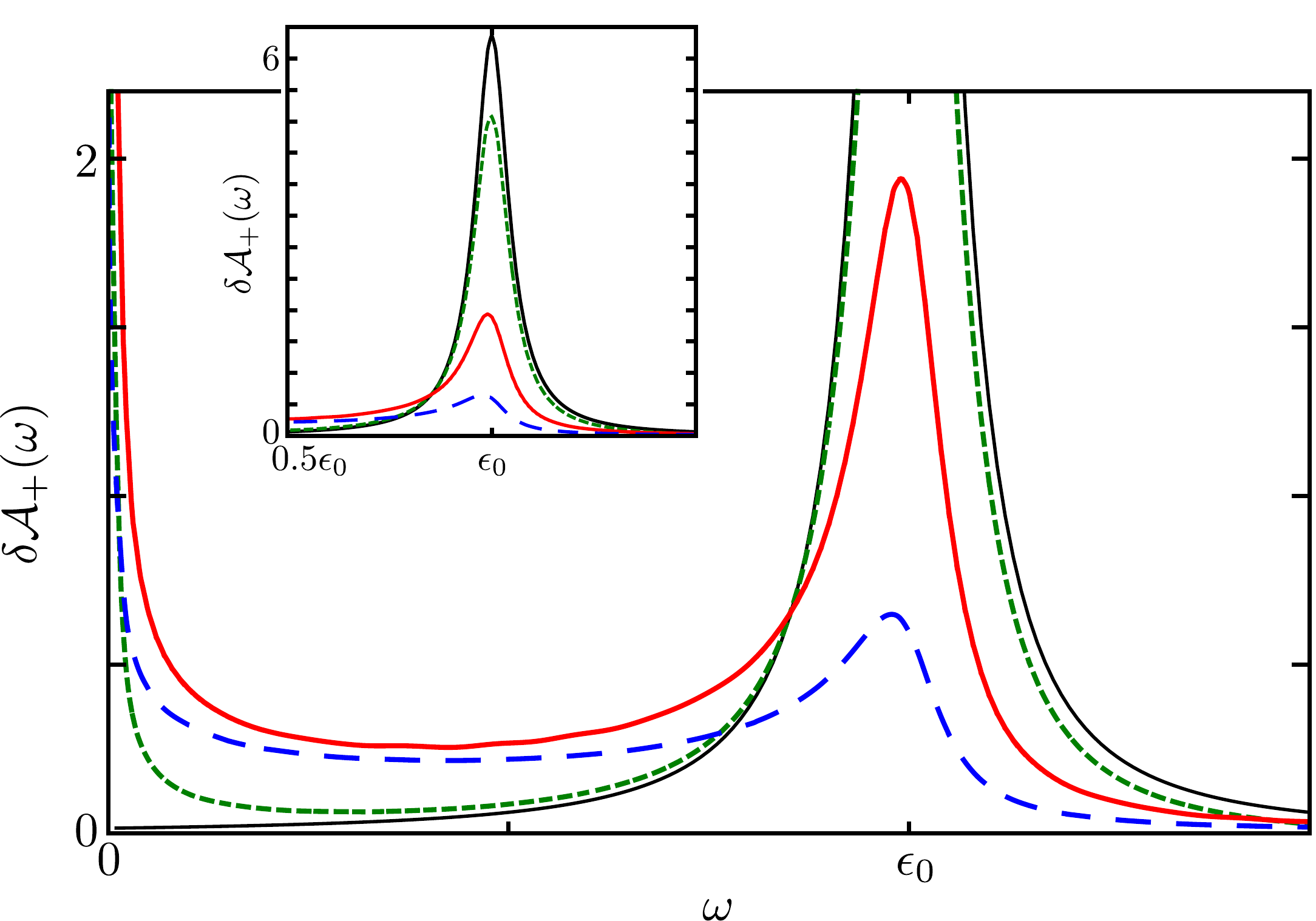}\\
	\includegraphics[width=\columnwidth]{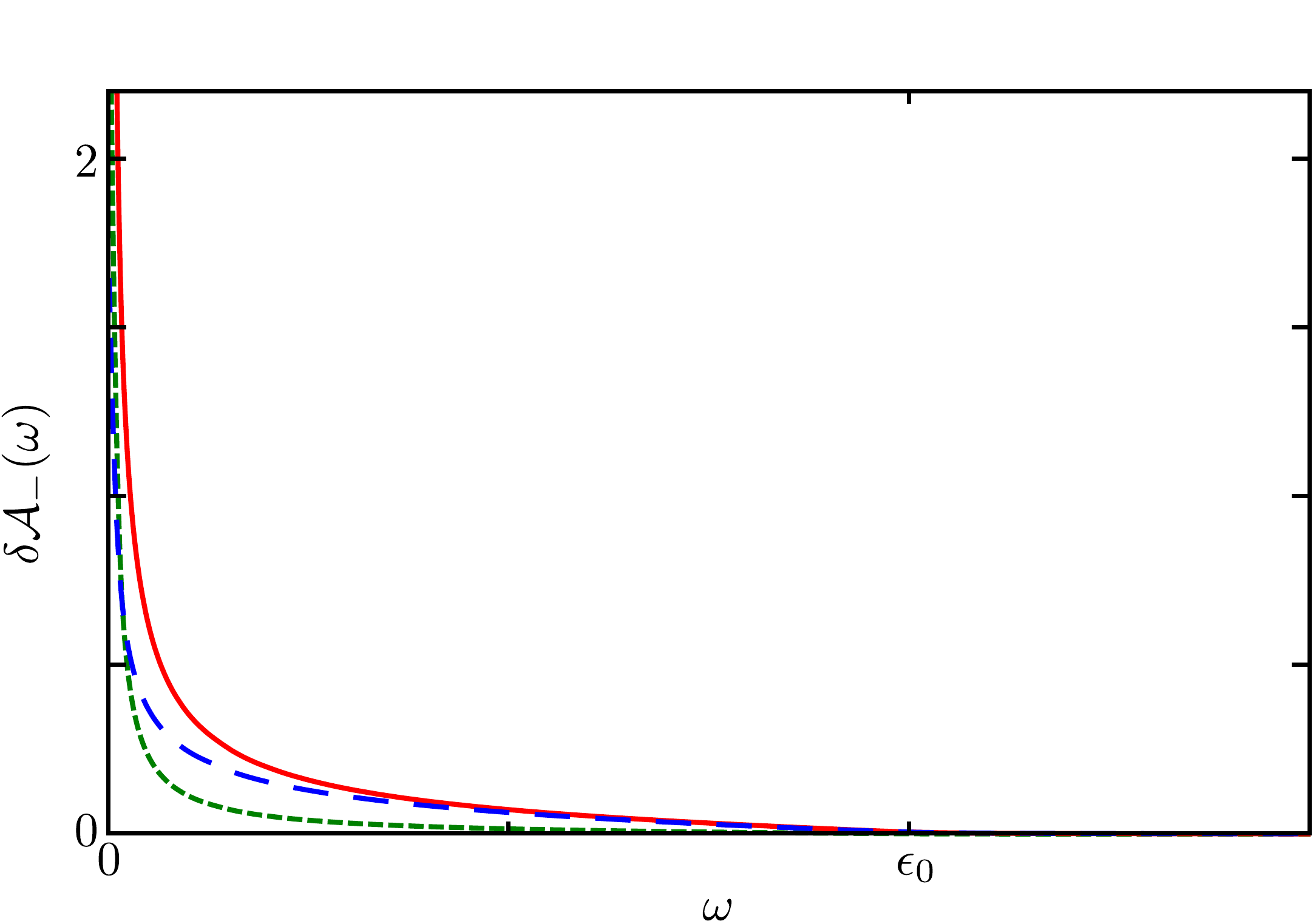}
	\caption{(color online) Total chiral energy distribution $\delta\mathcal{A}_\pm(\omega)$ (in units of $\epsilon_0^{-1}$) associated, respectively, with the chiral right-moving excitation $\delta \mathcal{A}_+$ (top panel) and the left-moving one $\delta \mathcal{A}_-$ (bottom panel). Different values of the interaction parameter $K$ are considered: $K=0.8$ (green short-dashed line), $0.54$ (red continuous line), $0.42$ (blue long-dashed line) and $1$ (thin black line). The inset in the top panel is a zoom of the peak centered around $\omega=\epsilon_0$. Parameters: $\epsilon_0au^{-1}=1/40$ and $\gamma=0.05\epsilon_0$.}
	\label{fig:energy-cfr}
\end{figure}

Fig.~\ref{fig:energy-cfr} shows the behavior of $\delta \mathcal{A}_\pm(\omega)$ obtained by using Eq.~\eqref{eq:A+-} and numerically evaluating Eq.~\eqref{eq:result-energy}. In analogy to the momentum distribution, the chiral right-moving component features a peak centered at $\omega=\epsilon_0$ (the average energy of the injected electron) which lowers and broadens as the interaction increases. However, in this case the broadening is highly asymmetric and tails increase only for energies $\omega<\epsilon_0$. This behavior is a consequence of energy conservation: on average, the total energy transferred to the LL by the electron injection is $\epsilon_0$ and it is therefore impossible to create more energetic excitations. Tails for $\omega>\epsilon_0$ are indeed just a consequence of the finite level broadening $\gamma$.
As the peak lowers, low-energy excitations appear near the Fermi energy both on $\delta \mathcal{A}_+$ (top panel) and $\delta \mathcal{A}_-$ (bottom panel), exhibiting a power law divergence at $\omega=0$. Indeed, in the limit $\omega\to 0^+$ the total chiral energy distributions read
\begin{equation}
\delta\mathcal{A}_\eta(\omega)= \frac{1}{\pi^2 \epsilon_0}\left(\frac{\epsilon_0 a}{u}\right)^{2A_-^2}\,\Gamma(1-2A_-^2)D\,\left(\frac{\omega}{\epsilon_0}\right)^{2A_-^2-1}\,,
\label{eq:scaling-energy}
\end{equation}
with
\begin{equation}
D=\sin^2(\pi A_+A_-)+\sin^2(\pi A_-^2)\,.
\end{equation}
Equation~\eqref{eq:scaling-energy} is demonstrated in Appendix~\ref{app:nescaling} and holds as long as $A_-^2<1/2$ ($K>0.27$). We observe that the divergence is integrable and features exactly the same exponents we already found in Eq.~\eqref{eq:scaling-nk} for the momentum distribution. {Once again, this exponent is robust with respect to the approximation in Eq.\ \eqref{eq:dot_correlator}.}

\begin{figure}
	\includegraphics[width=\columnwidth]{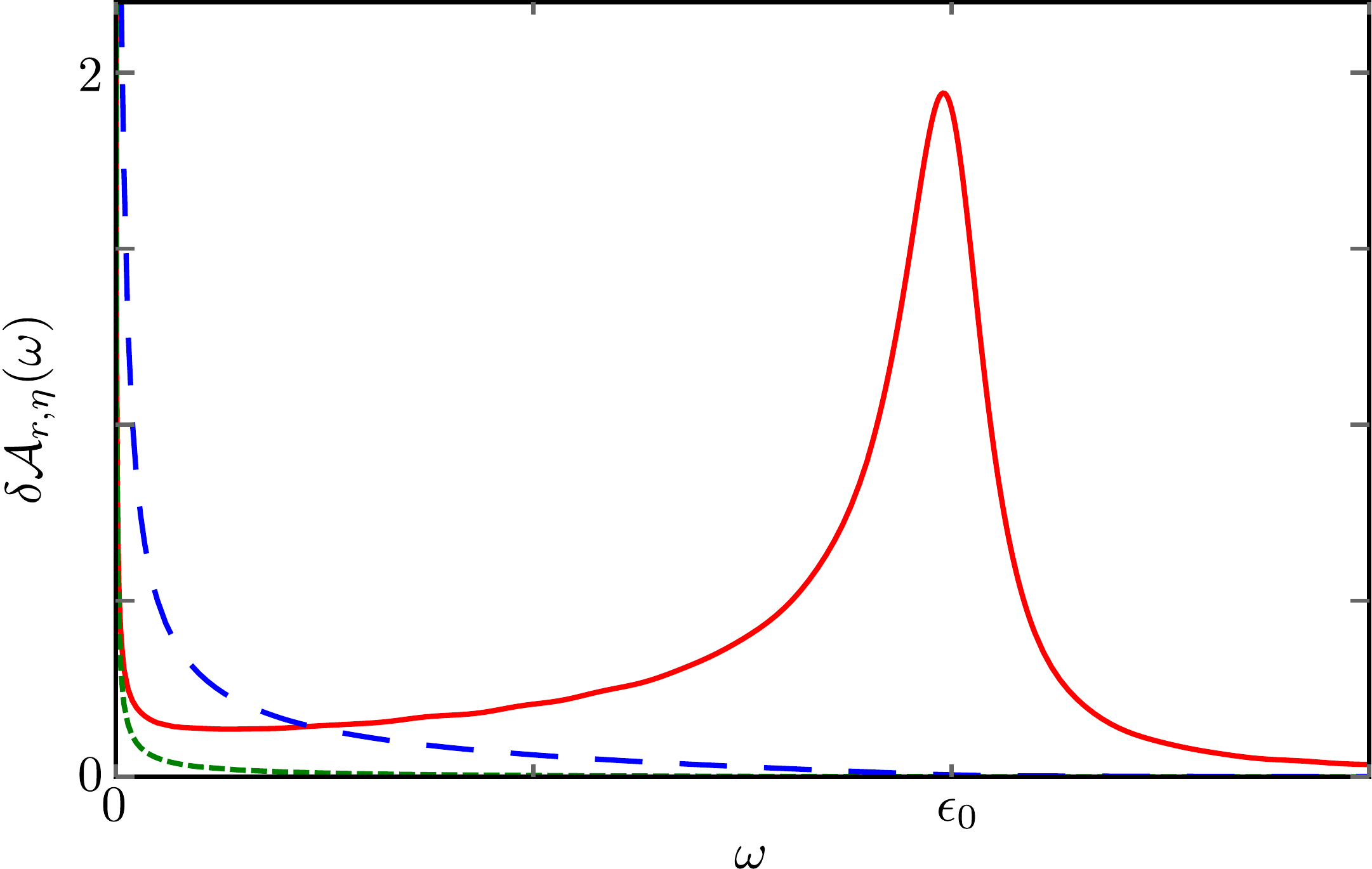}
	\caption{(color online) Chiral energy distributions $\delta \mathcal{A}_{r,\eta}(\omega)$ in units of $\epsilon_0^{-1}$. The solid red line refers to the $\delta \mathcal{A}_{R,+}$ contribution; the short-dashed green one refers to $\delta \mathcal{A}_{R,-}$. The long-dashed blue line refers to $\delta \mathcal{A}_{L,\pm}$. Parameters: $K=0.54$, $\epsilon_0au^{-1}=1/40$ and $\gamma=0.05\epsilon_0$.}
	\label{fig:energy}
\end{figure}

In Fig.~\ref{fig:energy} the contributions $\delta \mathcal{A}_{r,\eta}(\omega)$, associated with the $r=R$ and the $r=L$ branch for a given chirality $\eta$, are analyzed for a fixed interaction strength ($K=0.54$). The peak centered around $\epsilon_0$ is present only in $\delta \mathcal{A}_{R,+}$ (solid red line). Conversely, the majority of the low-energy excitations near the Fermi energy are hosted by the $L$ branch. In this respect, note that $\delta \mathcal{A}_{L,+}$ and $\delta \mathcal{A}_{L,-}$ coincide in the energy range we considered and they are both represented with the long-dashed blue line. Note that also $\delta \mathcal{A}_{L,\eta}(\omega)$ are strongly suppressed above $\epsilon_0$ as a consequence of energy conservation. As discussed for the momentum distribution, the creation of low-energy excitations on the $R$ branch comes from a higher-order process and it is thus less relevant. This can be clearly seen by observing the short-dashed green line representing $\delta A_{R,-}$ as well as the behavior of $\delta \mathcal{A}_{R,+}$ (solid red line) near the Fermi energy.

\section{Conclusion}\label{sec:conclusions}
In summary, we have discussed the fractionalization mechanism of a locally injected electron inside a Luttinger liquid. Due to the presence of $e$-$e$ interactions, two counterpropagating modes carrying a fractional charge start to form right after the injection. Long after the injection, these two modes are spatially separated, and we have studied the energy and momentum distributions in this situation.
The energy distribution provides information on how the energy of the injected electron is spread in the system. It features a peak near the energy of the injected electron, reminiscent of the initial Lorentzian distribution, which gets asymmetrically broadened and lowered as the interaction strength is increased. Correspondingly, part of the energy injected into the Luttinger liquid is redistributed by creating low-energy excitations, with a power law divergence as the energy approaches zero. Therefore, the injected single electron loses its single-particle nature by creating many-body excitations. The study of the momentum distribution allows for a detailed analysis of these excitations and distinguishes between particle and hole contributions. Interestingly, these are mostly excited within the channel not directly tunnel-coupled to the single-electron source. Moreover, the momentum distribution features a strong asymmetry between the two chiral fractional excitations, differently from their mirror-shaped profile in position space.

\begin{acknowledgments}
    GD and TLS acknowledge support by the National Research Fund Luxembourg (ATTRACT 7556175). MA, AC and MS acknowledge support by Research Fund of the University of Genova. 
\end{acknowledgments}

\appendix
\section{Calculation of the single electron coherence}\label{app:spc}
In this Appendix we give the details on the calculation of the single-electron coherence $\mathscr{G}_r(s,t;\xi,z)$ in Eq. \eqref{eq:spc}. The starting point is Eq.~\eqref{eq:general-average} together with the bosonization identity~\eqref{eq:bosonization}. Denote with $\hat O_r(s,t;\xi,z)$ the operator in the average~\eqref{eq:spc}. It has the property that $\hat O_r^\dagger(s,t;\xi,z)=\hat O_r(s,t;-\xi,-z)$. Then Eq.~\eqref{eq:g-gstar} immediately follows from Eq.~\eqref{eq:general-average} with
\begin{align}
g_r(s,t;\xi,z)&=|\lambda|^2\int_0^{t}\dif t_2\int_0^{t_2}\dif t_1\,e^{-\gamma(t_2+t_1)-i\epsilon_0(t_2-t_1)}\notag\\
&\times\mathscr{C}_r(t_1,s,t,\xi,z,t_2)\,,
\label{eq:app-spc:gr}
\end{align}
where $\mathscr{C}_r=\mathscr{C}_r^{(1)}-\mathscr{C}_r^{(2)}$ and
\begin{subequations}
	\begin{equation}
	\mathscr{C}_r^{(1)}=\Braket{\hat{\psi}_R(0,t_1)\hat O_r(s,t;\xi,z)\hat{\psi}^\dagger_R(0,t_2)}_\Omega\,,
	\end{equation}
	\begin{equation}
	\mathscr{C}_r^{(2)}=\Braket{\hat{\psi}_R(0,t_1)\hat{\psi}^\dagger_R(0,t_2)\hat O_r(s,t;\xi,z)}_\Omega\,.
	\end{equation}
\end{subequations}
Here, $\Braket{\dots}_\Omega$ denotes the ground state average. Let us focus on $\mathscr{C}_r^{(1)}$. First, the fermion fields are rewritten in terms of the chiral fields with the bosonization identity and Eq.~\eqref{eq:chiral-bosons}, so that the time evolution becomes chiral. Introducing the shorthand notations $x_\eta~=s-\eta ut$ and $\zeta_\eta=\xi-\eta uz$ we have:
\begin{widetext}
	\begin{align}
	\mathscr{C}_r^{(1)}&=\frac{e^{i\vartheta_rk_{\mathrm{F}}\xi}}{(2\pi a)^2}\prod_{\eta=\pm}\Braket{e^{-i\sqrt{2\pi}A_\eta\hat{\phi}_\eta(-\eta ut_1)}e^{i\sqrt{2\pi}A_{\vartheta_r\eta}\hat{\phi}_\eta(x_\eta-\zeta_\eta/2)}e^{-i\sqrt{2\pi}A_{\vartheta_r\eta}\hat{\phi}_\eta(x_\eta+\zeta_\eta/2)}e^{i\sqrt{2\pi}A_\eta\hat{\phi}_\eta(-\eta ut_2)}}_\Omega\label{eq:app-spc:average-c1}\\
	&=\frac{e^{i\vartheta_rk_{\mathrm{F}}\xi}}{(2\pi a)^2}\prod_{\eta=\pm}e^{2\pi A_\eta^2G_\eta(\eta u(t_2-t_1))}e^{2\pi A_{\vartheta_r\eta}^2G_\eta(-\zeta_\eta)}e^{2\pi A_\eta A_{\vartheta_r\eta}G_\eta(-\eta ut_1-x_\eta+\zeta_\eta/2)}e^{-2\pi A_\eta A_{\vartheta_r\eta}G_\eta(-\eta ut_1-x_\eta-\zeta_\eta/2)}\notag\\
	&\qquad\qquad\qquad\times e^{-2\pi A_\eta A_{\vartheta_r\eta}G_\eta(x_\eta-\zeta_\eta/2+\eta ut_2)}e^{2\pi A_\eta A_{\vartheta_r\eta}G_\eta(x_\eta+\zeta_\eta/2+\eta ut_2)}\,.
	\label{eq:app-spc:correlator1}
	\end{align}
\end{widetext}
The average~\eqref{eq:app-spc:average-c1} has been evaluated by using the identity\cite{vondelft}
\begin{equation}
\Braket{e^{\hat O_1}\dots e^{\hat O_n}}=e^{\frac{1}{2}\sum_{j=1}^n\braket{\hat O_j^2}}e^{\sum_{i<j}\braket{\hat O_i\hat O_j}}\,,
\end{equation}
with $n=4$, and the bosonic Green functions
\begin{align}
G_\eta(x)=\braket{\hat{\phi}_\eta(x)\hat{\phi}_\eta(0)}_\Omega-\braket{\hat{\phi}_\eta^2(0)}_\Omega=\frac{1}{2\pi}\ln\frac{a}{a-i\eta x}\,.
\label{eq:app-spc:green-functions}
\end{align}
Substituting Eq.~\eqref{eq:app-spc:green-functions} into Eq.~\eqref{eq:app-spc:correlator1} we find
\begin{align}
\mathscr{C}_r^{(1)}&=\frac{e^{i\vartheta_rk_{\mathrm{F}}\xi}}{(2\pi a)^2}\left[\frac{a}{a-iuz+i\vartheta_r\xi}\right]^{A_+^2}\left[\frac{a}{a-iuz-i\vartheta_r\xi}\right]^{A_-^2}\notag\\
&\;\;\times
\left[\frac{a}{a-iu(t_2-t_1)}\right]^{1+2 A_-^2}
\prod_{\eta=\pm}\Omega_{r,\eta}^{(1)}(x_\eta,\zeta_\eta,t_1,t_2)\,,
\label{eq:app-spc:result-c1}
\end{align}
	with
\begin{align}
\Omega_{r,\eta}^{(1)}(x_\eta,\zeta_\eta,t_1,t_2)&=\left[\frac{a+i\eta(x_\eta+\zeta_\eta/2+\eta ut_1)}{a+i\eta(x_\eta-\zeta_\eta/2+\eta ut_1)}\right]^{\alpha_{r,\eta}}\notag\\
&\times\left[\frac{a-i\eta(x_\eta-\zeta_\eta/2+\eta ut_2)}{a-i\eta(x_\eta+\zeta_\eta/2+\eta ut_2)}\right]^{\alpha_{r,\eta}}\,.
\label{eq:app-spc:omega}
\end{align}
The exponents $\alpha_{r,\eta}$ are defined in Eq. \eqref{eq:exponents}. The correlator $\mathscr{C}_r^{(2)}$ has the same structure as Eq.~\eqref{eq:app-spc:result-c1} but with different functions $\Omega_{r,\eta}^{(2)}$, which are readily obtained from $\Omega_{r,\eta}^{(1)}$ by taking the complex conjugate of all the factors containing $t_2$.

 The functions $\Omega_{r,+}^{(1,2)}$ and $\Omega_{r,-}^{(1,2)}$  correspond to the right and left moving packets respectively. Therefore their overlap becomes negligible in the long-time limit $t\gg\gamma^{-1}$ (i.e.\ when the injection is over).
 {In order to further clarify this point, let us focus on Eq.\ \eqref{eq:app-spc:omega}. The main features of functions $\Omega_{r,\eta}^{(1,2)}$, with respect to the $s$ variable, lie in the regions around their zeros and poles; everywhere else they are not significantly different from 1. If the time $t$ is much greater then all the other variables, the poles and zeros of $\Omega_{r,+}^{(1,2)}$ are well separated from those of $\Omega_{r,-}^{(1,2)}$ and the product in Eq.~\eqref{eq:app-spc:result-c1} can be thus converted into a sum }
 \begin{equation}
 \label{app:eq:prod}
 \prod_{\eta=\pm}\Omega_{r,\eta}^{(1,2)}(x_\eta,\zeta_\eta,t_1,t_2)=\sum_{\eta=\pm}\Omega_{r,\eta}^{(1,2)}(x_\eta,\zeta_\eta,t_1,t_2)-1\,.
 \end{equation}
 {A similar decomposition has been used also in Ref.\ [\onlinecite{sukhorukov2016prb}], where electron injection into interacting co-propagating channels is considered. Note that the condition $t\gg t_1,t_2$ is equivalent to $t\gg\gamma^{-1}$ because of the exponential suppression factor $e^{-\gamma(t_1+t_2)}$ present in \eqref{eq:app-spc:gr}. As for variables $\xi$ and $z$, a restriction of their integration domain such that they satisfy $|\xi|,|uz|\gg ut$ introduces uncertainties of the order of $(ut)^{-1}$ and $t^{-1}$ in the momentum and energy distribution respectively. In the long time limit one has $t\gg \gamma^{-1}\gg \epsilon_0^{-1}$ and these uncertainty thus become negligible.  Eq.\ \eqref{app:eq:prod} shows the separation of the two chiral contributions and the structure in Eq.~\eqref{eq:g-sum} is proven.} 
 
 Finally, the correlator $\mathscr{C}_r=\mathscr{C}_r^{(1)}-\mathscr{C}_r^{(2)}$ reads
\begin{align}
\mathscr{C}_r&=\frac{e^{i\vartheta_rk_{\mathrm{F}}\xi}}{(2\pi a)^2}\left[\frac{a}{a-iuz+i\vartheta_r\xi}\right]^{A_+^2}\left[\frac{a}{a-iuz-i\vartheta_r\xi}\right]^{A_-^2}\notag\\
&\times\left[\frac{a}{a-iu(t_2-t_1)}\right]^{1+2 A_-^2}\sum_{\eta=\pm}\Xi_{r,\eta}(x_\eta,\zeta_\eta,t_1,t_2)\,,
\label{eq:app-spc:cr}
\end{align}
where $\Xi_{r,\eta}=\Omega_{r,\eta}^{(1)}-\Omega_{r,\eta}^{(2)}$. In order to obtain  Eq.~\eqref{eq:g-r_eta-result} it is necessary to replace $t_2 = \tau+t_1$, performing the limit $t\to \infty$ and inserting the point-splitting term which is discussed in the next Appendix. 

\section{Point splitting procedure}\label{app:ps}
In this Appendix we discuss the point splitting procedure. As explained in the main text, this procedure results in the insertion of the multiplicative factor $(\zeta_\eta-ia\vartheta_r)\zeta_\eta^{-1}$ in the function $\mathcal{C}_{r,\eta}(\xi,z)$, see Eq.\ \eqref{eq:g-xi-z}. In the following we show that it ensures the correct representation of the excess particle density $\delta\rho_r$ in terms of the single electron coherence \begin{equation}
\delta \rho_r(s,t)= \delta\mathscr{G}_r(s,t;0,0)\,.
\label{eq:rho->G}
\end{equation}
We emphasize that this additional factor modifies the functions $g_{r,\eta}$ only near $\zeta_\eta=0$. Therefore the energy and momentum distribution will be affected by the point splitting procedure only for high energies/momenta, i.e. far away from the region we are interested in.

As shown in Ref.~[\onlinecite{calzona16energypart}], the excess particle density on the $r,\eta$ channel can be obtained by  computing the following bosonic expression directly
\begin{align}
&\delta\rho_{r,\eta}(s,t) \equiv -{\frac{\vartheta_r A_{\eta \vartheta_r}}{\sqrt{2	\pi}}} \, \delta \left[ \partial_s \phi_\eta(s-\eta ut )\right] \notag \\ 
&\quad =\eta\vartheta_r A_{\eta}A_{\eta\vartheta_r}\notag\\
&\quad\times\frac{|\lambda|^2}{\pi a}\mathrm{Re}\int_0^{t}\dif t_2\int_0^{t_2}\dif t_1 e^{-\gamma(t_1+t_2)-i\epsilon_0(t_2-t_1)}\notag\\
&\quad\times\left[\frac{a}{a-iu(t_2-t_1)}\right]^{1+2A_-^2}\frac{a/\pi}{a^2+[s-\eta u(t-t_2)]^2}\,.
\label{eq:app-ps:prb16}
\end{align}
This expression is consistent\cite{calzona16energypart} with the total injected charge given in Eq.\ \eqref{eq:sumrules}. Here, we show that the same result is obtained using the relation in Eq.\ \eqref{eq:rho->G} and the expressions summarized in Eqs. (\ref{eq:g-gstar}-\ref{eq:Xi}).  In fact, considering the limit of $g_{r,\eta}(s,t;\xi,z)$ for $(\xi,z)\to(0,0)$, a straightforward expansion of the functions $\Xi_{r,\eta}$ leads to
\begin{widetext}
	\begin{align}
	\lim_{\zeta_\eta\to 0}g_{r,\eta}(s,t;\zeta_\eta)&=\frac{|\lambda|^2}{(2\pi a)^2}\int_0^t\dif t_2\int_0^{t_2}\dif t_1 e^{-\gamma(t_1+t_2)}e^{-i\epsilon_0(t_2-t_1)}\left[\frac{a}{a-iu(t_2-t_1)}\right]^{1+2A_-^2}\notag\\
	&\quad\times\lim_{\zeta_\eta\to 0}\frac{\zeta_\eta-ia\vartheta_r}{\zeta_\eta}\left[\frac{2ia\eta\alpha_{r,\eta}\zeta_\eta}{a^2+(s-\eta u(t-t_2))^2}+\text{O}(\zeta_\eta^2)\right]=\notag\\
	&=\frac{|\lambda|^2}{2\pi a}\eta\vartheta_r\alpha_{r,\eta}\int_0^t\dif t_2\int_0^{t_2}\dif t_1 e^{-\gamma(t_1+t_2)}e^{-i\epsilon_0(t_2-t_1)}\left[\frac{a}{a-iu(t_2-t_1)}\right]^{1+2A_-^2}\frac{a/\pi}{a^2+[s-\eta u(t-t_2)]^2}\,.
	\end{align}
\end{widetext}
Then Eq.~\eqref{eq:app-ps:prb16} is recovered by taking into account the contribution of $g_r^*(s,t;-\zeta_\eta)$ and recalling that $\alpha_{r,\eta}=A_\eta A_{\eta\vartheta_r}$. Note that without the insertion of the point-splitting factor, the above limit would have been zero.

\section{Scaling of the momentum distribution}\label{app:nkscaling}
In this Appendix we derive the scaling behavior in Eq.~\eqref{eq:scaling-nk} of the momentum distribution near the Fermi points $\pm k_{\mathrm{F}}$. Four contributions need to be evaluated, but the calculation is very similar for each of them. First, we note that the behavior of functions $\delta n_{R/L,\eta}(k)$ for $k$ around $\pm k_{\mathrm{F}}$ is determined by large values of $\xi$ in the Fourier transform in Eq.\ \eqref{eq:result-nk}. Therefore we can safely neglect the cutoff $a$ with respect to $\xi$ as long as the integrals converge. In particular, for $A_-^2<1/2$, one has from Eq.\ \eqref{eq:g-xi-z}
\begin{equation}
\mathcal{C}_{r,\eta}(\xi,0)=\frac{a}{i\vartheta_r\xi}\left[\frac{a^2}{a^2+\xi^2}\right]^{A_-^2}\to\frac{a^{1+2 A_-^2}}{i\vartheta_r\xi}|\xi|^{-2 A_-^2}\,.
\label{eq:app-nkscaling:g}
\end{equation}
Let us now focus on $\delta n_{R,-}(k \approx k_\mathrm{F})$. The integral over the $s$ variable in Eq.~\eqref{eq:xi-nk} can be written, in the limit $a\to 0$, as
\begin{equation}
\int_{-\infty}^{+\infty}\chi_{R,-}(s,\xi,\tau)\dif s=-2i\xi\sin(\pi A_-^2){J}(\xi,\tau)\,,
\label{eq:app-nkscaling:xi-rm}
\end{equation}
with
\begin{align}
{J}(\xi,\tau)&=\int_0^1\dif x\left|\frac{x}{x-1}\frac{x-1-u\tau\xi^{-1}}{x-u\tau\xi^{-1}}\right|^{A_-^2}\notag\\
&\times\left[\theta(u\tau-\xi x)\theta(\xi x-\xi-u\tau)e^{i\pi A_-^2}\right.\notag\\
&\quad\left.+\theta(\xi x-u\tau)\theta(-\xi x+\xi+u\tau)e^{-i\pi A_-^2}\right]\,.
\label{eq:app-nkscaling:j}
\end{align}
Now, for large $\xi$ it is consistent to neglect $u\tau$ with respect to $\xi$, obtaining 
\begin{align}
J(\xi,\tau)\approx J_0(\xi)&=\theta(-\xi)e^{i\pi A_-^2}+\theta(\xi)e^{-i\pi A_-^2}\notag\\
&=e^{-i\pi A_-^2\,\mathrm{sgn}(\xi)}\,.
\label{eq:app-nkscaling:j0}
\end{align}
Inserting Eqs.~\eqref{eq:app-nkscaling:g}, \eqref{eq:app-nkscaling:xi-rm} and \eqref{eq:app-nkscaling:j0}  into Eq.~\eqref{eq:result-nk}, one finds 
\begin{align}
\delta n_{R,-}(k)&\approx-\frac{|\lambda|^2}{2\pi\gamma}\frac{a^{2 A_-^2}\sin(\pi A_-^2)}{2\pi^2a}\mathrm{Re}\left\{\,\int_0^{+\infty}\dif \tau\,\mathcal{F}(\tau)\right.\notag\\
&\times\left.\int_{-\infty}^{+\infty}\dif\xi\,\frac{e^{-i(k-k_{\mathrm{F}})\xi}}{|\xi|^{2 A_-^2}}e^{-i\pi A_-^2\,\mathrm{sgn}(\xi)}\right\}.
\end{align}
{Interestingly, the $\tau$-independence of $J_0(\xi)$ allows to compute the integral over $\xi$ without the need to know the function $\mathcal{F}(\tau)$}
\begin{align}
&\int_{-\infty}^{+\infty}\dif\xi\,\frac{e^{-i(k-k_\mathrm{F})\xi}}{|\xi|^{2 A_-^2}}e^{-i\pi A_-^2\,\mathrm{sgn}(\xi)}=\notag\\
&\qquad=2\Gamma(1-2A_-^2)\sin(2\pi A_-^2)|k-k_{\mathrm{F}}|^{2A_-^2-1}\theta(k_{\mathrm{F}}-k)\,.
\end{align}
{As a result, the power-law exponent $2A_-^2-1$ is robust with respect to the approximation made in Eq.\ \eqref{eq:dot_correlator}, which only affects the expression of $\mathcal{F}(\tau)$.}
	
The final step is to evaluate the real part of the integral over $\tau$. This can be done by exploiting the Fourier representation\cite{vannucci15,dolcetto14}
\begin{equation}
\left(\frac{a}{a-iu\tau}\right)^g=\frac{1}{\Gamma(g)}\left(\frac{a}{u}\right)^g\int_0^{+\infty}\dif E\,E^{g-1}e^{iE\tau}e^{-Ea/u}
\end{equation}
and leads to (for $\gamma\ll\epsilon_0$):\cite{calzona16energypart}
\begin{equation}
\mathrm{Re}\left\{\,\int_0^{+\infty}\dif\tau\,\mathcal{F}(\tau)\right\}=\frac{2\pi a\gamma}{|\lambda|^2}\,.
\label{eq:app-nkscaling:gamma-integral}
\end{equation}
Then one obtains
\begin{align}
\delta n_{R,-}(k\approx k_\mathrm{F})&\approx-\frac{a^{2A_-^2}}{\pi^2}\Gamma(1-2A_-^2)\sin(\pi A_-^2)\sin(2\pi A_-^2)\notag\\
&\quad\times|k-k_{\mathrm{F}}|^{2A_-^2-1}\theta(k_{\mathrm{F}}-k)\,.
\end{align}
The term $\delta n_{R,+}(k\approx k_\mathrm{F})$ follows in a similar way
\begin{align}
\delta n_{R,+}(k\approx k_\mathrm{F})&\approx\frac{a^{2A_-^2}}{\pi^2}\Gamma(1-2A_-^2)\sin(\pi A_-^2)\sin(2\pi A_-^2)\notag\\
&\quad\times|k-k_{\mathrm{F}}|^{2A_-^2-1}\theta(k-k_{\mathrm{F}})\,.
\end{align}
Thus the formula for $\delta n_R(k)$ in Eq.~\eqref{eq:scaling-nk} is proven by combining the last two equations. The calculation for $\delta n_L(k\approx-k_\mathrm{F})$ is similar to the one presented above, the only substantial difference being the exponents of the functions $\chi_{L,\eta}$, which are responsible for a different result when computing the integral over $\xi$. Let us consider for example the contribution $\delta n_{L,-}$. We have
\begin{equation}
\int_{-\infty}^{+\infty}\chi_{L,-}(s,\xi,\tau)\dif s=-2i\xi\sin(\pi A_+A_-)\bar{J}(\xi,\tau)\,,
\label{eq:app-nkscaling:xi-lm}
\end{equation}
where $\bar{J}$ is identical to Eq.~\eqref{eq:app-nkscaling:j} except for the exponent which now is $A_+A_-$ instead of $A_-^2$. Therefore, the asymptotic form of $\bar J(\xi, \tau)$ is $\bar{J}_0(\xi)=e^{-i\pi A_+A_-\text{sgn}(\xi)}$. Similar expressions hold for $\delta n_{L,+}$. Inserting these expressions, together with Eq.~\eqref{eq:app-nkscaling:g}, in Eq.~\eqref{eq:result-nk} one obtains
\begin{widetext}
	\begin{subequations}
		\begin{align}
		\delta n_{L,-}(k\approx-k_\mathrm{F})\approx&\,-\frac{a^{2A_-^2}}{\pi^2}\Gamma(1-2A_-^2)\sin(\pi A_+A_-)|k+k_\text{F}|^{2A_-^2-1}\notag\\
		&\times\{\sin[\pi(A_+A_-+ A_-^2)]\theta(-k-k_{\mathrm{F}})+\sin[\pi(A_-^2-A_+A_-)]\theta(k+k_\text{F})\}\,,
		\end{align}
		\begin{align}
		\delta n_{L,+}(k\approx-k_\mathrm{F})\approx&\,\frac{a^{2A_-^2}}{\pi^2}\Gamma(1-2A_-^2)\sin(\pi A_+A_-)|k+k_\text{F}|^{2A_-^2-1}\notag\\
		&\times\{\sin[\pi(A_+A_-+ A_-^2)]\theta(k+k_{\mathrm{F}})+\sin[\pi(A_-^2-A_+A_-)]\theta(-k-k_\text{F})\}\,.
		\end{align}
	\end{subequations}
\end{widetext}
The formula for $\delta n_L(k)$ in Eq.~\eqref{eq:scaling-nk} follows from the last two expressions.

\section{Scaling of the energy distribution}\label{app:nescaling}
In this Appendix we evaluate the expression in Eq.~\eqref{eq:scaling-energy} of the energy distribution for small $\omega$. In particular, we focus on the term $\delta\mathcal{A}_{R,-}$ in  Eq.~\eqref{eq:result-energy}. The other contributions can be evaluated in the same way. First of all, we note that the behavior of function $\delta \mathcal{A}_{R,-}$ at small $\omega$ is described by large $z$ in the Fourier transform in Eq.\ \eqref{eq:result-energy}. One then has 
\begin{equation}
\mathcal{C}_{R,-}(0,z)=\left[\frac{a}{a-iuz}\right]^{1+2A_-^2}\to\frac{ia^{1+2A_-^2}}{zu^{1+2A_-^2}}\frac{e^{i\pi A_-^2\mathrm{sign}(z)}}{|z|^{2A_-^2}}\,.
\end{equation}
Moreover, in complete analogy with the previous Appendix, the integral over $s$ in Eq.\ \eqref{eq:result-energy} can be expressed as
\begin{equation}
\int_{-\infty}^{+\infty}\chi_{R,-}(s,uz,\tau)\dif s=-2iuz\sin(\pi A_-^2)J(uz,\tau)\,,
\end{equation}
where function $J$ and its asymptotic expression are provided in Eqs.\ \eqref{eq:app-nkscaling:j} and\ \eqref{eq:app-nkscaling:j0} respectively. Next,  substituting these expressions in Eq.~\eqref{eq:result-energy}, one obtains
\begin{align}
\delta \mathcal{A}_{R,-}(\omega)&\approx\frac{|\lambda|^2}{2\pi\gamma}\frac{\sin(\pi A_-^2)}{2\pi^2a}\mathrm{Re}\left\{\,\int_0^{+\infty}\dif \tau\,\mathcal{F}(\tau)\right.\notag\\
&\left.\times\left(\frac{a}{u}\right)^{2 A_-^2}\int_{-\infty}^{+\infty}\dif z\,\frac{e^{i\omega z}}{|z|^{2 A_-^2}}\right\}\,.
\end{align}
The integral over $z$ yields
\begin{equation}
\int_{-\infty}^{+\infty}\dif z\,\frac{e^{i\omega z}}{|z|^{2 A_-^2}}=2\Gamma(1-2A_-^2)\sin(\pi A_-^2)\omega^{2A_-^2-1}\,.
\end{equation}
Finally, using Eq.~\eqref{eq:app-nkscaling:gamma-integral}, one finds ($\omega>0$)
\begin{equation}
\delta\mathcal{A}_{R,-}(\omega \approx 0)\approx\left(\frac{a}{u}\right)^{2A_-^2}\frac{\Gamma(1-2A_-^2)}{\pi^2}\sin^2(\pi A_-^2)\omega^{2A_-^2-1}\,.
\end{equation}
The result for $\delta\mathcal{A}_{R,+}(\omega)$ is exactly the same, while for the $L$-channel we also find $\delta\mathcal{A}_{L,+}(\omega)=\delta\mathcal{A}_{L,-}(\omega)$, with
\begin{equation}
\delta\mathcal{A}_{L,-}(\omega \approx 0)\approx\left(\frac{a}{u}\right)^{2A_-^2}\frac{\Gamma(1-2A_-^2)}{\pi^2}\sin^2(\pi A_+A_-)\omega^{2A_-^2-1}\,.
\end{equation}
Combining the last two results we readily arrive at Eq.~\eqref{eq:scaling-energy}.


\begin{thebibliography}{43}
	\bibitem{grenier2011electronoptics} C. Grenier, R. Herv\'{e}, G. F\`{e}ve, and P. Degiovanni, Mod. Phys. Lett. B {\bf 25}, 1053 (2011).
	\bibitem{bocquillon2014electron}E. Bocquillon, V. Freulon, F.D. Parmentier, J.-M Berroir, B. Pla\c{c}ais, C. Wahl, J. Rech, T. Jonckheere, T. Martin, C. Grenier, D. Ferraro, P. Degiovanni, and G. F\`{e}ve, Annalen der Physik {\bf 526}, 1 (2014).
	\bibitem{henny99-hbt} M. Henny, S. Oberholzer, C. Strunk, T. Heinzel, K. Ensslin, M. Holland, and C. Sch\"onenberger, Science {\bf 284}, 296 (1999).
	\bibitem{oliver99-hbt} W. D. Oliver, J. Kim, R. C. Liu, and Y. Yamamoto, Science {\bf 284}, 299 (1999).
	\bibitem{jin03-mz}Y. Ji, Y. Chung, D. Sprinzak, M. Heiblum, D. Mahalu, and H. Shtrikman, Nature {\bf 422}, 415 (2003).
	\bibitem{feve2007mesoscopic}G. F\`{e}ve, A. Mah\'{e}, J.-M.Berroir, T.Kontos, B.Pla\c{c}ais, D.C. Glattli, A. Cavanna, B. Etienne, and Y. Jin, Science {\bf316}, 1169 (2007).
	\bibitem{mahe2010mesoscopic} A. Mah\'{e}, F. D. Parmentier, E. Bocquillon, J.-M. Berroir, D. C. Glattli, T. Kontos, B. Pla\c{c}ais, G. F\`{e}ve, A. Cavanna, and Y. Jin, Phys. Rev. B {\bf82}, 201309 (2010).
	\bibitem{buttiker1993mesoscopic}M. B\"{u}ttiker, H. Thomas, and A. Pr\^{e}tre, Phys. Lett. A {\bf180}, 364 (1993).
	\bibitem{moskalet2008mesoscopic} M. Moskalets, P. Samuelsson, and M. B\"{u}ttiker, Phys. Rev. Lett. {\bf100},086601 (2008).
	\bibitem{levitov96-97-06} L. S. Levitov, H. Lee and G. B. Lesovik, J. Math. Phys. {\bf 37}, 4845 (1996); D. A. Ivanov, H. W. Lee, and L. S. Levitov, Phys. Rev B {\bf 56}, 6839 (1997); J. Keeling, I. Klich, and L. S. Levitov, Phys. Rev. Lett. {\bf 97}, 116403 (2006).
	\bibitem{rech2017} J. Rech, D. Ferraro, T. Jonckheere, L. Vannucci, M. Sassetti, and T. Martin, Phys. Rev. Lett. {\bf118}, 076801 (2017).
	\bibitem{dubois2013levitonsNature} J. Dubois, T. Jullien, F. Portier, P. Roche, A. Cavanna, Y. Jin, W. Wegscheider, P. Roulleau, and D. C. Glattli, Nature (London) {\bf 502}, 659 (2013).
	\bibitem{hasan2010colloquium}M. Z. Hasan and C. L. Kane, Rev. Mod. Phsy. {\bf 82}, 3045 (2010).
	\bibitem{qi2011topological}X.-L. Qi and S.-C. Zhang, Rev. Mod. Phys. {\bf 83}, 1057 (2011).
	\bibitem{bhz2006} B. A. Bernevig, T. L. Hughes, and S.-C. Zhang, Science {\bf314}, 1757 (2006).
	\bibitem{dolcettoreview} G. Dolcetto, M. Sassetti, and T. Schmidt, Riv. Nuovo Cimento {\bf 39}, 113 (2016).
	\bibitem{konig2007quantum}M. Koenig, S. Videmann, C. Brune, A. Roth, H. Buhmann, L. W. Molenkamp, X.-L. Qi, and S.-C. Zhang, Science {\bf 318}, 766 (2007).
	\bibitem{liu2008inasgasb} C. C. Liu, T. L. Hughes, X.-L. Qi, K. Wang, and S.-C. Zhang, Phys. Rev. Lett. {\bf100}, 236601 (2008).
	\bibitem{lingjie2015inasgasb} L. Du, I. Knez, G. Sullivan, and R.-R. Du, Phys. Rev. Lett. {\bf114}, 096802 (2015).
	\bibitem{knez2011inasgasb} I. Knez, R.-R.Du, and G. Sullivan, Phys.Rev.Lett. {\bf107}, 136603 (2011).
	\bibitem{giamarchi2003book} T. Giamarchi, \textit{Quantum Physics in One Dimension} (Oxford University Press, Oxford, 2003).
	\bibitem{landau-fermiliquid}L. D. Landau, Sov. Phys. JETP {\bf 3}, 920 (1957); {\em ibid.} {\bf 5}, 101 (1957); {\em ibid.} {\bf 8}, 70 (1959).
	\bibitem{tomonaga1950} S.-I. Tomonaga, Prog. Theor. Phys. {\bf 5}, 544 (1950).
	\bibitem{luttinger1963} J. M. Luttinger, J. Math. Phys. {\bf 4}, 1154 (1963).
	\bibitem{haldane86-ll} F. D. M. Haldane, J. Phys. C {\bf 14}, 2585 (1981).
	\bibitem{auslaender02-spincharge}O. M. Auslaender, A. Yacoby, R. de Picciotto, K. W. Baldwin, L. N. Pfeiffer, and K. W. West, Science {\bf 295}, 825 (2002).
	\bibitem{jompol09-spincharge}Y. Jompol, C. J. B. Ford, J. P. Griffiths, I. Farrer, G. A. C. Jones, D. Anderson, D. A. Ritchie, T. W. Silk, and A. J. Schofield, Science {\bf 325}, 597 (2009).
    \bibitem{schmidt09} T. L. Schmidt, A. Imambekov, and L. I. Glazman, Phys. Rev. Lett. {\bf 104}, 116403 (2010), T. L. Schmidt, A. Imambekov, and L. I. Glazman, Phys. Rev. B {\bf 82}, 245104 (2010).
	\bibitem{kamata17-spincharge}M. Hashisaka, N. Hiyama, T. Akiho, K. Muraki, T. Fujisawa, Nat. Phys. {\bf 13}, 559 (2017).
	\bibitem{deshpande2010electron} V. V. Deshpande, M. Bockrath, L. I. Glazman, and A. Yacoby, Nature {\bf 464}, 209 (2010).
	\bibitem{barak2010interacting} G. Barak, H. Steinberg, L. N. Pfeiffer, K. W. West, L. Glazman, F. Von Oppen, and A. Yacoby, Nature Physics {\bf 6}, 489 (2010).
	\bibitem{maslov1995landauer}D. L. Maslov and M. Stone, Phys. Rev. B {\bf 52}, 5539(R) (1995).
	\bibitem{safi1995transport}I. Safi and H.J.  Schulz, Phys. Rev. B {\bf 52}, 17040(R) (1995).
	\bibitem{garate2012noninvasive}I. Garate and K. Le Hur, Phys. Rev. B {\bf 85}, 195465 (2012).
	\bibitem{calzona2015physe}A. Calzona, M. Carrega, G. Dolcetto, and M. Sassetti, Physica E 74, 630 (2015).
	\bibitem{muller17} T. M\"{u}ller, R. Thomale, B. Trauzettel, E. Bocquillon, and O.  Kashuba, Phys. Rev. B {\bf 95}, 245114 (2017). 
	\bibitem{safi1997properties} I. Safi,  Ann. Phys. (France) {\bf 22}, 463 (1997).
	\bibitem{steinberg2007charge} H. Steinberg, G. Barak, A. Yacoby, L. N. Pfeiffer, K. W. West, B. I. Halperin, and K. Le Hur, Nature Physics {\bf 4}, 116 (2007).
	\bibitem{kamata2014fractionalized}H. Kamata, N. Kumada, M. Hashisaka, K. Muraki, and T. Fujisawa, Nature NanoTech. {\bf 9}, 177 (2014).
	\bibitem{perfetto2014time}E. Perfetto, G. Stefanucci, H. Kamata, and T. Fujisawa, Phys. Rev B {\bf 89}, 201413(R) (2014).
	\bibitem{pham2000fractional} K.-V. Pham, M. Gabay, and P. Lederer, Phys. Rev. B {\bf 61}, 16397 (2000).
	\bibitem{dolcetto13prb} G. Dolcetto, N. Traverso Ziani, M. Biggio, F. Cavaliere, and M. Sassetti, Phys. Rev. B {\bf 87}, 235423 (2013).
	\bibitem{calzona2015spin} A. Calzona, M. Carrega, G. Dolcetto, and M. Sassetti, Phys. Rev. B {\bf 92}, 195414 (2015).
	\bibitem{das2011} S. Das and S. Rao, Phys. Rev. Lett. {\bf 106}, 236403 (2011). 
	\bibitem{karzig2011energypart} T. Karzig, G. Refael, L. I. Glazman, and F. von Oppen, Phys. Rev. Lett. {\bf 107}, 176403 (2011).
	\bibitem{calzona16energypart} A. Calzona, M. Acciai, M.Carrega, F.Cavaliere, and M. Sassetti, Phys. Rev. B {\bf 94}, 035404 (2016).
	\bibitem{calzona2017quench-frac} A. Calzona, F. M. Gambetta, M. Carrega, F. Cavaliere, and M. Sassetti, Phys. Rev. B {\bf 95} 085101 (2017); A. Calzona, F. M. Gambetta, F. Cavaliere, M. Carrega, and M. Sassetti, arXiv:1706.01676.
	\bibitem{bocquillon2012electron} E. Bocquillon, F. D. Parmentier, C. Grenier, J.-M. Berroir, P. Degiovanni, D. C. Glattli, B. Pla\c{c}ais, A. Cavanna, Y. Jin, and G. F\`{e}ve, Phys. Rev. Lett. {\bf108}, 196803 (2012).
	\bibitem{bocquillon2013hom} E. Bocquillon, V. Freulon, J.-M. Berroir, P. Degiovanni, B. Pla\c{c}ais, A.Cavanna,Y.Jin, and G. F\`{e}ve, Science {\bf339}, 1054 (2013).
	\bibitem{jullien14} T. Jullien, P. Roulleau, B. Roche, A. Cavanna, Y. Jin, and D. C. Glattli, Nature {\bf 514}, 603 (2014).
	\bibitem{fevenatcomm}
	V. Freulon, A. Marguerite, J.-M. Berroir, B. Pla\c{c}ais, A. Cavanna, Y. Jin, and G. F\`{e}ve, Nat. Comm. {\bf 6}, 6854 (2015).
	\bibitem{tewari16}S. Tewari, P. Roulleau, C. Grenier, F. Portier, A. Cavanna, U. Gennser, D. Mailly, and P. Roche, Phys. Rev. B {\bf 93}, 035420 (2016).
	\bibitem{jonckeere12hom}T. Jonckheere, J. Rech, C. Wahl, and T. Martin,
	Phys. Rev. B {\bf 86}, 125425 (2012).
	\bibitem{wahl14}C. Wahl, J. Rech, T. Jonckheere, and T. Martin, Phys. Rev. Lett. {\bf 112}, 046802 (2014).
	\bibitem{degiovanni09relaxation}P. Degiovanni, C. Grenier, and G. F\`eve, Phys. Rev. B {\bf 80}, 241307(R) (2009).
	\bibitem{lunde10relaxation}A. M. Lunde, S. E. Nigg, and M. B\"{u}ttiker, Phys. Rev. B {\bf 81}, 041311(R) (2010).
	\bibitem{degiovanni10relaxation}P. Degiovanni, C. Grenier, G. F\`eve, C. Altimiras, H. le Sueur, and F. Pierre, Phys. Rev. B {\bf 81}, 121302(R) (2010).
	\bibitem{levkivskj12relaxation}I. P. Levkivskyi and E. V. Sukhorukov
	Phys. Rev. B {\bf 85}, 075309 (2012).
	\bibitem{ferraro14decoherence}D. Ferraro, B. Roussel, C. Cabart, E. Thibierge, G. F\`eve, C. Grenier, and P. Degiovanni, Phys. Rev. Lett. {\bf 113}, 166403 (2014); A. Marguerite, C. Cabart, C. Wahl, B. Roussel, V. Freulon, D. Ferraro, Ch. Grenier, J.-M. Berroir, B. Pla\c cais, T. Jonckheere, J. Rech, T. Martin, P. Degiovanni, A. Cavanna, Y. Jin, and G. F\`eve, Phys. Rev. B {\bf 94}, 115311 (2016).
	\bibitem{sukhorukov2016prb} A. O. Slobodeniuk, E. G. Idrisov, and E. V. Sukhorukov, Phys. Rev. B {\bf 93}, 035421 (2016).
	\bibitem{sueur10relaxation}H. le Sueur, C. Altimiras, U. Gennser, A. Cavanna, D. Mailly, and F. Pierre, Phys. Rev. Lett. {\bf 105}, 056803 (2010).
	\bibitem{altimiras10relaxation} C. Altimiras, H. le Sueur, U. Jennser, A. Cavanna, D. Mailly, and F. Pierre, Nat. Phys. {\bf 6}, 34 (2010).
	\bibitem{inhofer2013} A. Inhofer and D. Bercioux, Phys. Rev. B {\bf88}, 235412 (2013).
	\bibitem{hofer2013} P. P. Hofer, and M. B\"{u}ttiker, Phys. Rev. B {\bf88}, 241308 (2013).
	\bibitem{ferraro14homtopo} D. Ferraro, C. Wahl, J. Rech, T. Jonckheere, and T. Martin, Phys. Rev. B {\bf 89}, 075407 (2014).
	\bibitem{strom2015entaglemntspin} A. Str\"{o}m, H. Johannesson, and P. Recher, Phys. Rev. B {\bf91}, 245406 (2015).
	\bibitem{dolcetto16entanglement} G. Dolcetto and T. L. Schmidt, Phys. Rev. B {\bf 94}, 075444 (2016).
	\bibitem{dolcini11-16} F. Dolcini, Phys. Rev. B {\bf 83}, 165304 (2011); F. Dolcini, Phys. Rev. B {\bf 85}, 033306 (2012); F. Dolcini, R. C. Iotti, A. Montorsi, and F. Rossi, Phys. Rev. B {\bf 94}, 165412 (2016).
	\bibitem{Li15} T. Li, P. Wang, H. Fu, L. Du, K. A. Schreiber, X. Mu, X. Liu, G. Sullivan, G. A. Cs\'athy, X. Lin, and R.-R. Du, Phys. Rev. Lett. {\bf115}, 136804 (2015).
	\bibitem{Wu06} C. Wu, B. A. Bernevig, and S.-c. Zhang, Phys. Rev. Lett. {\bf 96}, 106401 (2006). 
	\bibitem{vondelft} J. von Delft, and  H. Schoeller, Ann. Phys. {\bf7 }, 225 (1998).
	\bibitem{voit} J. Voit, Rep. Prog. Phys. {\bf 58}, 977 (1995).
	\bibitem{miranda} E. Miranda, Braz. J. Phys. {\bf 33}, 3 (2003).
	\bibitem{kleimann02} T. Kleimann, F. Cavaliere, M. Sassetti, B. Kramer, Phys. Rev. B {\bf 66}, 165311 (2002).
	\bibitem{chernii14} I. Chernii, I. P. Levkivskyi, and E. V. Sukhorukov, Phys. Rev. B {\bf 90}, 245123 (2014).
	\bibitem{elste10} F. Elste, D. R. Reichman, and A. J. Millis, Phys. Rev. B {\bf 81}, 205413 (2010).
	\bibitem{iyoda2014} E. Iyoda, T. Kato, K. Koshino, and T. Martin, Phys. Rev. B {89}, 205318 (2014).
	\bibitem{note_correlator} {It is worth noting that, in the non-interacting case, it is possible to solve the problem at all orders in lambda without the need of such an approximation\cite{vasseur13}.}
	\bibitem{vasseur13} Romain Vasseur, Kien Trinh, Stephan Haas, and Hubert Saleur
	Phys. Rev. Lett. {\bf 110}, 240601 (2013).
	\bibitem{wachter07} P. W\"{a}chter, V. Meden, and K. Sch\"{o}nhammer,  Phys. Rev. B {\bf 76}, 125316 (2007).
	\bibitem{lerner08} I. V. Lerner, V. I. Yudson, and I. V. Yurkevich,  Phys. Rev. Lett. {\bf 100}, 256805 (2008).	
	\bibitem{grenier11} C. Grenier, R. Herv\'{e}, E. Bocquillon, F.D. Parmentier, J.-M Berroir, G. F\`{e}ve, and P. Degiovanni, New J. Phys. {\bf 13}, 093007 (2011).
	\bibitem{glauber62-63} R. Glauber, Phys. Rev. Lett. {\bf 10}, 84 (1962); R. Glauber, Phys. Rev. {\bf 130}, 2529 (1963); R. Glauber, \emph{ibid.}\ {\bf 131}, 2766 (1963).
	\bibitem{note_pointsplitting} The insertion of the point-splitting factor only affects functions $g_{r,\eta}$ around the point $z=\xi=0$. It will not change significantly the momentum and energy distribution in the region we are interested in.
	\bibitem{ferraro13wigner} D. Ferraro, A. Feller, A. Ghibaudo, E. Thibierge, E. Bocquillon, G. F\`eve, C. Grenier, and P. Degiovanni, Phys. Rev. B {\bf 88}, 205303 (2013).
    \bibitem{anderson67} P. W. Anderson, Phys. Rev. Lett. {\bf 18}, 1049 (1967).
    \bibitem{mahan81} G. D. Mahan, {\it Many-body Physics} (Plenum, New York, 1981).
	\bibitem{voit93spectral} J. Voit, J. Phys.: Cond. Mat. {\bf 5}, 8305 (1993).
	\bibitem{takei2010spectroscopy} S. Takei, M. Milletar\`i, and B. Rosenow, Phys. Rev. B {\bf 82}, 041306(R) (2010).
	\bibitem{meden1992} V. Meden and K.	Schonhammer,  Phys. Rev. B {\bf 46}, 15753 (1992).
	\bibitem{vannucci15} L. Vannucci, F. Ronetti, G. Dolcetto, M. Carrega, and M. Sassetti, Phys. Rev. B {\bf 92}, 075446 (2015).
	\bibitem{dolcetto14} G. Dolcetto, F. Cavaliere, and M. Sassetti, Phys. Rev. B {\bf 89}, 125419 (2014).
\end{thebibliography}
\end{document}